\documentstyle[pra,aps]{revtex}
\begin{document}
\draft 
\input epsf

\def\simlt{\stackrel{<}{{}_\sim}}
\def\simgt{\stackrel{>}{{}_\sim}}
\def\sfrac#1#2{{\textstyle{\sst{#1}\over\sst{#2}}}}
\newcommand{\Frac}[2]{\frac{{\displaystyle #1}}{{\displaystyle #2}}}

\title{Largest temperature of the radiation era\\
and its cosmological implications}

\author{Gian Francesco Giudice\thanks{On leave of absence from INFN, 
Sezione di Padova, Padua, Italy.}
}
\address{CERN Theory Division, CH-1211 Geneva 23, Switzerland}
     
\author{Edward W.\ Kolb
}
\address{NASA/Fermilab Astrophysics Center, Fermi
     National Accelerator Laboratory, Batavia, Illinois \  60510-0500,\\
     and Department of Astronomy and Astrophysics, Enrico Fermi Institute, \\
     The University of Chicago, Chicago, Illinois \ 60637-1433}

\author{Antonio Riotto
}
\address{Scuola Normale Superiore, Piazza dei Cavalieri 7, 
     I-56126 Pisa, Italy,\\
     and INFN, Sezione di Pisa, I-56127 Pisa, Italy}

\date{May 2000}

\maketitle

\begin{abstract}
The thermal history of the universe before the epoch of
nucleosynthesis is unknown. The maximum temperature in the
radiation-dominated era, which we will refer to as the reheat
temperature, may have been as low as 0.7 MeV.  In this paper we show
that a low reheat temperature has important implications for many
topics in cosmology.  We show that weakly interacting massive
particles (WIMPs) may be produced even if the reheat temperature is
much smaller than the freeze-out temperature of the WIMP, and that the
dependence of the present abundance on the mass and the annihilation
cross section of the WIMP differs drastically from familiar results.
We revisit predictions of the relic abundance and resulting model
constraints of supersymmetric dark matter, axions, massive neutrinos,
and other dark matter candidates, nucleosynthesis constraints on
decaying particles, and leptogenesis by decay of superheavy particles.
We find that the allowed parameter space of supersymmetric models is
altered, removing the usual bounds on the mass spectrum; the
cosmological bound on massive neutrinos is drastically changed, ruling
out Dirac (Majorana) neutrino masses $m_\nu$ only in the range 
33 keV $\simlt m_\nu\simlt$ 6 (5) MeV, which is significantly
smaller from the the standard disallowed range 94 eV $\simlt
m_\nu\simlt$ 2 GeV (this implies that massive neutrinos may still play
the role of either warm or cold dark matter); the cosmological upper
bound on the Peccei-Quinn scale may be significantly increased to $
10^{16}$GeV from the usually cited limit of about
$10^{12}$GeV; and that efficient out-of-equilibrium GUT baryogenesis
and/or leptogenesis can take place even if the reheat temperature is
much smaller than the mass of the decaying superheavy particle.
\end{abstract}

\pacs{PACS: 98.80.Cq \hfill 
hep-ph/0005123 \hfill   SNS-PH/00-05 \hfill FNAL-Pub-00/075-A\hfill 
CERN-TH/2000-107}

 \section{Introduction} \label{intro}

The initiation of the radiation-dominated era of the universe is
believed to result from the decay of coherent oscillations of a scalar
field whose energy dominated the universe before decay.  The decay of
the coherent oscillations of the scalar field and the subsequent
thermalization of the decay products is known as
reheating.\footnote{While we discuss reheating as the decay of
coherent field oscillations, we only make use of the fact that the
energy density of coherent field oscillations scales in expansion as
$a^{-3}$, where $a$ is the Friedmann--Robertson--Walker scale
factor. One could just as easily imagine that the universe is
dominated by some unstable massive particle species, rather than
coherent oscillations of a scalar field.}

The reheat process is often associated with the final stage of
inflation.  However, reheating could have been episodic, with several
reheat events after inflation. We will be interested in the final
reheating before primordial nucleosynthesis, which may just as well
have been the result of the decay of a weakly coupled scalar field
unrelated to inflation, for instance a modulus. For this reason the
scalar field $\phi$, whose decay leads to reheating, will not be
referred to as the inflaton.

A common assumption is that many of the interesting cosmological
phenomena accessible to present-day observations occurred after
reheating, during the radiation-dominated phase of the early universe.
This seems to be a reasonable assumption, since inflation
\cite{revinf} erases any initial condition on the number densities of
ordinary particles, while the reheat process repopulates the
universe. The assumption of an initial condition of 
thermal and chemical equilibrium in a radiation-dominated universe is
then equivalent to the hypothesis that the maximum temperature
obtained during the radiation-dominated era, $T_{RH}$, is larger than
characteristic temperatures of cosmological processes under
investigation.

The fact that we have no physical evidence of the radiation-dominated
era before the epoch of nucleosynthesis ({\it i.e.,} temperatures
above about 1 MeV) is a simple, but crucial, point.  Therefore, {\it a
priori} one should consider $T_{RH}$ as an unknown quantity that can
take any value as low as about 1 MeV.  Indeed, there are good physical
motivations for studying cosmologies with very low $T_{RH}$.

In theories in which the weakness of gravity is explained by large
compactified extra dimensions \cite{qg}, emission of Kaluza-Klein
gravitons in the bulk constrains the normalcy temperature at which the
radius of the compactified dimensions is stabilized with vanishing
energy density in the compactified space to be in the MeV to GeV
region \cite{Arkani-Hamed:1999nn}.  In practice, the stabilization
occurs in a post-inflationary phase, and therefore we can
identify the normalcy temperature with $T_{RH}$.

Supergravity and superstring theories usually have particles, such as
a gravitino or a modulus, with only gravitational
interactions.  Late decay of these particles may jeopardize the 
success of standard big-bang nucleosynthesis
\cite{lindley}. This problem can be solved by assuming sufficiently
low $T_{RH}$, of the order of $10^8$ to $10^{10}$ GeV \cite{gravitino}.
Moreover, it has been recently realized \cite{gra} that nonthermal
production of these gravitational relics during the inflationary phase
can impose upper bounds on $T_{RH}$ as low as 100 GeV.

In this paper, we will show that the phenomenological point of view
that the reheat temperature may be as low as 1 MeV has rich
implications for particle dark matter, neutrino mass limits, axion
cosmology, and baryogenesis.

The key point of our considerations is that reheating is not an
instantaneous process. On the contrary, the radiation-dominated phase
follows a prolonged stage of matter domination during which the energy
density of the universe is dominated by the coherent oscillations of
the field $\phi$.  The oscillations start at time $H_I^{-1}$ and end
when the age of the universe becomes of order of the lifetime
$\Gamma_\phi^{-1}$ of the scalar field. At times $H_I^{-1} \simlt t
\simlt \Gamma_\phi^{-1}$ the dynamics of the system is quite
involved. During this stage the energy density per comoving volume of
the $\phi$ field decreases as $\exp(-\Gamma_\phi t)$ and the light
decay products of the scalar field thermalize.  The temperature $T$ of
this hot plasma, however, does not scale as $T \propto a^{-1}$ as in
the ordinary radiation-dominated phase ($a$ is the
Friedmann--Robertson--Walker scale factor) \cite{ckr,book,pr}, but
reaches a maximum $T_{\rm MAX} \sim (H_I M_{Pl})^{1/4}T_{RH}^{1/2}$
($M_{Pl}$ is the Planck mass)
and then decreases as $T \propto a^{-3/8}$, signalling the continuous
release of entropy from the decays of the scalar field. This scaling
continues until the time $t \sim \Gamma_\phi^{-1}$ when the
radiation-dominated phase commences with temperature
$T_{RH}$. Therefore, before reheating is completed, for a given
temperature the universe expands faster than in the
radiation-dominated phase.  Notice that $T_{RH}$ is not the maximum
temperature during the reheat process. On the contrary, $T_{\rm MAX}$
can be much larger than $T_{RH}$.  The behavior of the universe during
reheating is discussed in detail in Sec.\ \ref{reheating}.

In Sec.\ \ref{relic} we will use these results to compute the relic
abundance of a dark-matter species ($X$) produced during reheating.
We will consider the case that $T_{RH}$ is smaller than the $X$
freeze-out temperature.\footnote{For a different perspective, see
Ref.\ \cite{nontherm}.}  Although naively one might expect a negligible 
$X$ number density under these circumstances, we find that the $X$
relic density can reach cosmologically interesting values.  We also
show that because of entropy release during the reheat stage, for a
given mass and cross section the present $X$ abundance is smaller than
obtained assuming freeze out in the radiation-dominated era.  This
relaxes the bounds coming from requiring that $\Omega_Xh^2\simlt 1$.
Moreover, the parametric dependence on the mass and the annihilation
cross section of the present abundance is nonstandard. 
Therefore, most of the cosmological constraints on specific
particle properties have to be revisited.

The value of the maximum temperature during reheating, $T_{\rm MAX}$,
delineates different regions for our results.  If $T_{\rm MAX}$ is
smaller than the $X$ mass, $X$ particles generated by collisions in
the thermal bath during the coherent $\phi$ oscillations are always
nonrelativistic. If the cross section is small enough, the $X$
particles do not reach chemical equilibrium and the present abundance
is {\it proportional} to the cross section, $\Omega_X\propto\langle
\sigma_A |v|\rangle$, in contrast with the usual
radiation-dominated case in which $\Omega_X\propto \langle
\sigma_A |v|\rangle^{-1}$.  Avoiding overclosure of the universe
imposes a {\it lower} bound on the annihilation cross section. On the
other hand, if the $X$ particles reach chemical equilibrium before
reheating is completed, the ratio of the $X$ number density to entropy
density in a comoving volume does not stop decreasing when the
particles freeze out because entropy is released until reheating is
over. Therefore, the present abundance does not depend only on
$\langle \sigma_A |v|\rangle^{-1}$ (as in the standard case), but also
on the reheat temperature $T_{RH}$: the lower $T_{RH}$, the smaller
the abundance.  We will provide a formula for $\Omega_X$ which
reproduces both the standard result when $T_{RH}$ is equal to the
freeze-out temperature of the relic particle and the nonstandard
result $\Omega_X\propto\langle \sigma_A |v|\rangle$ when the value
cross section becomes smaller than some critical value.

If $T_{\rm MAX}$ and $T_{RH}$ are larger than the $X$ mass, we will
show that the relevant processes determining the present $X$ abundance
occur during reheating or afterwards.

An important result is that $T_{\rm MAX}$ (or equivalently $H_I$) is
relevant when deciding if $X$ is relativistic or not, but does not
appear in the final expression for the relic abundance. Therefore, once
$T_{\rm MAX}$ has determined the pertinent case, $\Omega_X$ depends on
the physics of the $\phi$ field only through $T_{RH}$.  It is easy to
understand why. The $X$ number density results from the competition of
two rates, the interaction rate and the expansion rate of the
universe. Before reheating is completed the expansion rate depends
only on $T_{RH}$, $H\sim T^4/T^2_{RH} M_{Pl}$, and therefore the
final abundance depends only upon $T_{RH}$.

In Sec.\ \ref{applications} we discuss the applications of our
findings to some popular cold dark matter candidates.  Here, we
preview some of our results.

While excluded in the usual scenario, thermal WIMPs with mass larger
than the unitarity bound of a few hundred TeV \cite{kam} may be viable
dark matter candidates in low reheat models.

Of the many WIMP candidates, the best motivated seems to be the
neutralino, the lightest supersymmetric particle (LSP)
\cite{dmreview}. In the case in which the LSP is mainly a Bino,
requiring $\Omega_{B}\simlt 1$ gives an upper bound on the slepton
mass $\tilde{m}_{\ell_R}$ \cite{bees,kane,falk}. However,
once we relax the assumption that the reheat temperature is higher
than the freeze-out temperature of the WIMP, we will show that the
upper bound on $\tilde{m}_{\ell_R}$ is drastically relaxed and may
completely disappear. The same argument can be applied to 
other supersymmetric candidates.

Another striking application for dark matter that illustrates our
point is the computation of the relic abundance of massive neutrinos.
The well-known cosmological Cowsik--McClelland---Lee--Weinberg bound
\cite{cw,lw} rules out neutrinos more massive than roughly 90 eV and
lighter than around 2 GeV. This result has a significant impact in
cosmology, ruling out, for instance, the possibility that neutrinos
are warm dark matter. This standard result, however, assumes that the
reheat temperature is much higher than 1 MeV and that neutrinos have
been in chemical equilibrium up to temperatures of the order of 1 MeV.
We will show that if the reheat temperature is as small as allowed by
big-bang nucleosynthesis, then massive, stable Dirac (Majorana) 
neutrinos are compatible with cosmology if they are lighter than 
about 33 keV or heavier than about 6 (5) MeV. This implies, for instance, 
that neutrinos may still be warm or cold dark matter and play a 
significant role in other cosmological or astrophysical phenomena.

We then proceed by investigating the implications of a low reheat
temperature for axion cosmology. It is well known that in the standard
scenario the oscillations of the axion field generated by the
misalignment mechanism overclose the universe unless the Peccei-Quinn
scale, $f_{PQ}$, is smaller than about $10^{12}$ GeV. This bound,
however, is obtained assuming that the reheat temperature is larger
than the QCD scale. When this assumption is abandoned, the
cosmological upper bound on $f_{PQ}$ is significantly relaxed to
$f_{PQ}\simlt 10^{16}$ GeV if $T_{RH}\sim 1$ MeV. 

If the $X$ particle has a nonvanishing but small decay lifetime
$\tau_X$, its decay products may destroy the light elements generated
during primordial nucleosynthesis.  This gives strong constraints in
the plane $(M_X,\tau_X)$ \cite{sarkar} that are very sensitive to the
number density of the species $X$ at freeze out. Lowering the reheat
temperature implies a smaller number density and therefore much less
restrictive bounds. 

These results all imply that presently stated cosmological limits may
not always be relevant in limiting particle properties such as the
supersymmetric mass spectra in the experimentally verifiable range of
future colliders.

As a last application, in Sec.\ \ref{baryolepto} we analyze the
production of unstable superheavy states during the process of
reheating, keeping in mind the possibility that the subsequent decay of
these states may generate the observed baryon asymmetry. The fact that
$T_{\rm MAX}$ is larger than the reheat temperature may give rise to
an efficient production of these superheavy states.  As a result,
out-of-equilibrium GUT baryogenesis and/or leptogenesis can take place
even if the reheat temperature is much smaller than the mass of the
superheavy decaying particle. This is particularly useful in
supersymmetric scenarios where $T_{RH}$ has to be low enough to avoid
the overproduction of gravitinos and other dangerous relics.

Finally, Sec.\ \ref{conclusionsect} summarizes our results.

\section{The Dynamics of Reheating}  \label{reheating}

\subsection{The Relevant Boltzmann Equations}  \label{boltzmann}

In this section we study the Boltzmann equations for the time
evolution of a system whose energy density is in the form of unstable
massive particles $\phi$, stable massive particles $X$, and radiation
$R$ (other similar studies can be found in Refs.\ 
\cite{ckr,book,pr,jmd,ther1,ther2}).
We assume that $\phi$ decays into radiation with a rate $\Gamma_\phi$,
and that the $X$ particles are created and annihilate into radiation
with a thermal-averaged cross section times velocity $\langle \sigma v
\rangle$. The corresponding energy and number densities satisfy the
differential equations \cite{ckr}
\begin{eqnarray}
\frac{d \rho_\phi}{dt} &=& -3H \rho_\phi-\Gamma_\phi \rho_\phi
\label{binp} \\
\frac{d \rho_R}{dt} &=& -4H \rho_R +\Gamma_\phi \rho_\phi +
\langle \sigma v \rangle 2\langle E_X \rangle 
\left[ n_X^2 - \left({n_X^{eq}}\right)^2\right]
\label{binr} \\
\frac{d n_X}{dt} &=& -3H n_X
-\langle \sigma v \rangle  \left[ n_X^2 -
\left({n_X^{eq}}\right)^2\right] 
\label{binx} \ .
\end{eqnarray}
Here, we assume that each $X$ has energy $\langle E_X \rangle \simeq
\sqrt{M^2 + 9T^2}$ and the factor $2\langle E_X \rangle$ is the
average energy released in $X$ annihilation.  Later we will assume
$\rho_X=\langle E_X \rangle n_X$.  The Hubble expansion parameter $H$
is given by
\begin{equation}
H^2=\frac{8\pi}{3M_{Pl}^2} (\rho_\phi +\rho_R +\rho_X) \ .
\end{equation}
The equilibrium number density for particles obeying Maxwell-Boltzmann
statistics can be expressed in terms of $K_2$, the modified Bessel
function of the second kind:
\begin{eqnarray}
n_X^{eq} & = & \frac{g T^3}{2\pi^2} \left(\frac{M_X}{T}\right)^2K_2(M_X/T) 
\nonumber \\
& \longrightarrow & \Frac{gT^3}{\pi^2} \quad (T\gg M) \nonumber \\ 
& \longrightarrow & g \left(\Frac{M_XT}{2\pi}\right)^{3/2}\exp(-M_X/T)
\quad (T\ll M) \ , 
\end{eqnarray}
where $g$ is the number of degrees of freedom of the $X$-particle species.

For cosmological considerations it is more appropriate to express
$\Gamma_\phi$ in terms of the reheat temperature $T_{RH}$ using the
conventional expression 
\begin{equation}
\Gamma_\phi =\sqrt{\frac{4\pi^3 g_*(T_{RH})}{45}} \frac{T_{RH}^2}{M_{Pl}}  \ ,
\end{equation}
where $g_*(T)$ describes the effective number of degrees of freedom at
temperature $T$.  This expression defines $T_{RH}$.

Next, we express Eqs.\ (\ref{binp})--(\ref{binx}) in terms of
dimensionless variables and convert time derivatives to derivatives
with respect to the scale factor $a$.  The dimensionless variables we
choose are
\begin{equation}
\Phi \equiv \rho_\phi T_{RH}^{-1}a^3 \ ; \quad R\equiv \rho_R a^4 \ ; 
\quad X \equiv n_X a^3 \ ; 
\quad A \equiv a/a_I \ .
\end{equation}  
The choice of $T_{RH}^{-1}$ in the definition of $\Phi$ is for
convenience; any mass scale would suffice.  The factor $a_I$ will be
chosen as the initial value of the scale factor for the integration.
Since no physical result can depend upon the choice of $a_I$, 
we are free to choose
\begin{equation}
a_I=T_{RH}^{-1}\ . 
\end{equation}  
In terms of the new variables, Eqs.\ (\ref{binp})--(\ref{binx}) become
\begin{eqnarray}
\frac{d\Phi}{dA} &=& - \left(\frac{\pi^2g_*}{30}\right)^{1/2} \! \!
        \frac{A^{1/2}\Phi}{\sqrt{\Phi+R/A+X\langle E_X \rangle /T_{RH}}}
\label{butp} \\
\frac{dR}{dA} &=&  \left(\frac{\pi^2g_*}{30}\right)^{1/2} \! \!
        \frac{A^{3/2}\Phi}{\sqrt{\Phi+R/A+X\langle E_X \rangle /T_{RH}}}
+ \left(\frac{3}{8\pi}\right)^{1/2} \! \!
         \frac{A^{-3/2} \langle\sigma v\rangle 2\langle E_X\rangle M_{Pl} }
         {\sqrt{\Phi+R/A+X\langle E_X \rangle /T_{RH}}} 
         \left( X^2-X_{eq}^2\right)
\label{butr} \\
\frac{dX}{dA} &=& - \left(\frac{3}{8\pi}\right)^{1/2} \! \!
\frac{A^{-5/2}  \langle\sigma v\rangle  M_{Pl} T_{RH}} 
        {\sqrt{\Phi+R/A+X\langle E_X \rangle /T_{RH}}} 
        \left(X^2-X_{eq}^2\right) \ . 
\label{butx}
\end{eqnarray}

At early times the energy density of the universe is completely
dominated by the $\phi$ field. The initial value of the $\phi$ energy
density can be expressed in terms of the initial expansion rate,
$H_I$, as $\rho_\phi = (3/8\pi)M_{Pl}^2H_I^2$.  Therefore, we will
solve Eqs.\ (\ref{butp})--(\ref{butx}) choosing the following initial
conditions:
\begin{equation}
\Phi_I =\frac{3}{8\pi}\frac{M_{Pl}^2 H_I^2}{T_{RH}^4} \ , 
\quad R_I=X_I=0 \ , \quad A_I=1 \ .
\label{init}
\end{equation}

\subsection{The Temperature--Scale Factor Relation} 
\label{tempscale}

During the epoch between the initial time, $H_I^{-1}$, and the
completion of reheating at time $\Gamma_\phi^{-1}$, the temperature of
the universe does not scale as $T\sim a^{-1}$ as in the
radiation-dominated era, but follows a different law \cite{ckr}. This
unusual relation between the temperature and the scale factor, derived
below, will significantly affect the calculation of the relic
abundance of $X$ particles.

The temperature of the system is measured by the radiation energy
density, and therefore $T$ is related to $R$ by
\begin{equation}
T=\left[ \frac{30}{\pi^2 g_*(T)}\right]^{1/4} \frac{R^{1/4}}{A} T_{RH} \ .
\label{terr}
\end{equation}
At early times ($H\gg \Gamma_\phi$), we can approximate the right-hand
side of Eq.\ (\ref{butr}) by retaining only the terms proportional to
the $\phi$ energy density and by taking $\Phi \simeq \Phi_I$. The
solution of Eq.\ (\ref{butr}) then becomes
\begin{equation}
R \simeq \frac{2}{5} \left(\frac{\pi^2g_*}{30}\right)^{1/2} 
\left( A^{5/2} -1 \right) \Phi_I^{1/2} \ .
\end{equation}

Using this result in Eq.\ (\ref{terr}) we obtain the expression
of the temperature $T$ as a function of the scale factor,
\begin{equation}
T=T_{\rm MAX}f(A) \ ,
\label{tempgen}
\end{equation}
where $T_{\rm MAX}$ and $f(A)$ are given by 
\begin{eqnarray}
T_{\rm MAX} & \equiv &
\left( \frac{3}{8}\right)^{2/5} \left( \frac{5}{\pi^3}\right)^{1/8}
\frac{g_*^{1/8}(T_{RH})}{g_*^{1/4}(T_{\rm MAX})} M_{Pl}^{1/4} H_I^{1/4}
T_{RH}^{1/2} \nonumber \\
 & = &\left[ \frac{g_*(T_{RH})}{10}\right]^{1/8} 
\left[ \frac{10}{g_*(T_{\rm MAX})}\right]^{1/4}
\left( \frac{H_I}{\rm eV}\right)^{1/4}
\left( \frac{T_{RH}}{\rm 100\, MeV}\right)^{1/2}\, 42\,{\rm GeV}\ ,\nonumber \\
f(A) & \equiv & \kappa
        \left( A^{-3/2}-A^{-4} \right)^{1/4} \ .
\end{eqnarray}
The constant $\kappa$ is defined as
\begin{equation}
\kappa   \equiv  \left(\frac{8^8}{3^35^5}\right)^{1/20}
        \frac{g_*^{1/4}(T_{\rm MAX})}{g_*^{1/4}(T)} =
1.3 \, \frac{g_*^{1/4}(T_{\rm MAX})}{g_*^{1/4}(T)}    \ .
\end{equation}
The function $f(A)$ starts as zero, then grows until
$A_0=(8/3)^{2/5}$, where it reaches its maximum $f(A_0)=1$
(corresponding to $T=T_{\rm MAX}$), and then decreases as
$A^{-3/8}$. Therefore, for $A>A_0$, Eq.\ (\ref{tempgen}) can be
approximated by
\begin{equation}
T\simeq  \kappa T_{\rm MAX} A^{-3/8}=\left[ \frac{9g_*(T_{RH})}
{5\pi^3g^2_*(T)}\right]^{1/8}  M_{Pl}^{1/4} H_I^{1/4} T_{RH}^{1/2} A^{-3/8} \ .
\label{temp}
\end{equation}
This result shows that during the phase before reheating the
temperature has a less steep dependence on the scale factor than in
the radiation-dominated era.  In other words, as the temperature
decreases, the universe expands faster before reheating than in the
radiation-dominated epoch.  Notice also that $T_{\rm MAX}$ can be much
larger than $T_{RH}$, as long as $H_I>T_{RH}^2/M_{Pl}$.

\subsection{The Temperature--Expansion Rate Relation} 
\label{hscale}

Next, consider the temperature dependence of the expansion rate $H$
during the epoch of reheating.  Between the time when $T_{\rm MAX}$ is
obtained and the decay time $\Gamma_\phi^{-1}$, the scalar field
energy density scales as $\rho_\phi=\Phi_I T_{RH}^4 A^{-3}$.  Since
$H^2\simeq (8\pi/3) \rho_\phi/M_{Pl}^2$, we can express $H$ as
\begin{equation}
H^2 = \frac{8\pi}{3} \frac{\Phi_IT_{RH}^4 A^{-3}}{M_{Pl}^2} \ . 
\end{equation}
Now we can use Eq.\ (\ref{temp}) to express $A$ in terms of $T$, with
the result
\begin{equation}
H = \left[ \frac{5\pi^3g^2_*(T)}{9g_*(T_{RH})}\right]^{1/2} 
\frac{T^4}{T_{RH}^2M_{Pl}} \ .
\label{hhhh}
\end{equation}

This result can be compared to the result for a radiation-dominated
universe $(H \propto T^2)$ and a matter-dominated universe $(H
\propto T^{3/2})$.

\section{Calculation of the Relic Abundance}   \label{relic}

In this paper we are interested in the situation in which the $X$
particles never obtain chemical equilibrium in the radiation-dominated
era after reheating.\footnote{Here we make the usual distinction
between chemical equilibrium and kinetic equilibrium.  Kinetic
equilibrium can be achieved by $X$-number conserving scatterings, such
as $\gamma X \longleftrightarrow \gamma X$ ($\gamma$ represents a
light degree of freedom like a photon). Chemical equilibrium can only
be achieved by processes that change the number of $X$ particles, such
as $XX \longleftrightarrow \gamma \gamma$. For massive particles the
cross section for the second process may be orders of magnitude
smaller than the cross section for the first process, and it is
possible to assume kinetic equilibrium but not chemical equilibrium. }
This means that $T_{RH}$ must be smaller than the conventional
freeze-out temperature (roughly equal to $M_X/20$ for nonrelativistic,
weakly interacting particles). In this situation we can encounter
several possibilities.  In the first case, the $X$ particles are
always nonrelativistic and never in chemical equilibrium, either
before or after reheating. In the second case, the nonrelativistic $X$
particles reach chemical equilibrium, but then freeze out before the
completion of the reheat process.  Finally, we can consider the case
when the relevant processes of particle production and freeze out
occur before reheating at temperatures at which $X$ is still
relativistic. In this section, we will compute the $X$ thermal relic
abundance in all these cases.

In principle, the relic abundance could receive contributions from
other sources, like the direct $\phi$ decay into $X$ particles
\cite{rm}, or from the production and decay of heavy particles
eventually decaying into $X$. In this paper we will ignore these
model-dependent 
effects and therefore our calculation can be viewed as a lower bound
on the $X$ abundance.

The nonrelativistic and relativistic cases are discriminated by the
conditions $M_X>T_{\rm MAX}$ and $M_X<T_{\rm MAX}$, respectively. This
translates into a condition on $H_I$; for instance, the
nonrelativistic case corresponds to
\begin{equation}
H_I  <  \left( \frac{8}{3}\right)^{8/5} \left( \frac{\pi^3}{5}\right)^{1/2}
\frac{g_*(T_{\rm MAX})}{g_*^{1/2}(T_{RH})}\frac{M_X^4}{M_{Pl}T_{RH}^2} =
\left[ \frac{g_*(T_{\rm MAX})}{10}\right] 
\left[ \frac{10}{g_*(T_{RH})}\right]^{1/2}
\left( \frac{M_X}{\rm 100\,GeV}\right)^{4}
\left( \frac{\rm 100\,MeV}{T_{RH}}\right)^{2}\, 31\,{\rm eV} \ .
\label{ineq}
\end{equation}

In this paper we will treat $T_{RH}$ and $T_{\rm MAX}$ as free
parameters and we will not rely on particular models of inflation or
$\phi$ decay. Nevertheless, it is useful to show what kind of $\phi$
physics can give rise to the different cases considered in this
section. First of all, we are interested in very low reheat
temperatures. This can be achieved if $\phi$ has a typical
gravitational decay width $\Gamma_\phi \sim M_\phi^3/M_{Pl}^2$, for
which
\begin{equation}
T_{RH} \sim \left( \frac{10}{g_*}\right)^{1/4}
\left( \frac{M_\phi}{100\, {\rm TeV}}\right)^{3/2}\, 4\, {\rm MeV}.
\end{equation}
The value of $T_{\rm MAX}$ is determined by the initial $\phi$ energy
density.  If $M_\phi$ is the mass scale characterizing the physics of
$\phi$, we can expect $\rho_\phi(a_I) \sim M_\phi^4$.  This happens, for
instance, in hybrid models of inflation.  In this case we find
\begin{equation}
T_{\rm MAX} \sim\left( \frac{10}{g_*}\right)^{1/4}
\left( \frac{M_\phi}{100\, {\rm TeV}}\right)^{5/4}\, 10\,{\rm GeV}.
\end{equation}
On the other hand, in the case of chaotic inflation, one expects that
the $\phi$ field has an initial value of the order of the Planck mass
and therefore $\rho_\phi(a_I) \sim M_\phi^2 M_{Pl}^2$. This leads to
\begin{equation}
T_{\rm MAX} \sim\left( \frac{10}{g_*}\right)^{1/4}
\left( \frac{M_\phi}{100\, {\rm TeV}}\right)\,30\, {\rm TeV}.
\end{equation}
For an $X$ particle with typical electroweak mass and for 
$M_\phi \sim 100$ TeV, the two options
correspond to the nonrelativistic and relativistic case,
respectively.

\subsection{Nonrelativistic Nonequilibrium Production and Freeze Out}  
\label{nrnoneq}

Let us suppose that the $X$ particles are always nonrelativistic and
the condition in Eq.\ (\ref{ineq}) is satisfied.  Since we are
considering the case in which $X$ does not reach chemical equilibrium
($X\ll X_{eq}$), at early times, Eq.\ (\ref{butx}) can be approximated
by
\begin{equation}
\frac{dX}{dA} = {\left(\frac{3}{8\pi}\right)^{1/2}A^{-5/2} \, 
\langle\sigma v\rangle  M_{Pl} T_{RH} \, X_{eq}^2} \, \Phi_I^{-1/2} \ . 
\label{butxx}
\end{equation}
The equilibrium distribution in the nonrelativistic limit is
\begin{equation}
X_{eq}=gA^3  \left(\frac{M_XT}{2\pi T_{RH}^2}\right)^{3/2} \exp(-M_X/T) \ ,
\label{eqnox}
\end{equation}
where $g$ is the number of degrees of freedom of the particle $X$ and
the temperature $T$ is given by Eq.\ (\ref{temp}). We express the
thermal-averaged annihilation cross section times velocity as
\begin{equation}
\langle \sigma v \rangle \equiv \frac{1}{M_X^2} 
\left( \alpha_s+\frac{T}{M_X}\alpha_p
\right) .
\label{cross}
\end{equation}
Here the dimensionless coefficients $\alpha_s$ and $\alpha_p$ describe,
respectively, the $s$-wave and $p$-wave annihilations in a
nonrelativistic expansion of the cross section. Using Eqs.\
(\ref{init}), (\ref{temp}), (\ref{eqnox}), and (\ref{cross}) in Eq.\
(\ref{butxx}), we obtain
\begin{equation}
\frac{dX}{dA}= \frac{g^2}{(2\pi)^3} 
\frac{M_X\kappa^3 T_{\rm MAX}^3}{H_IT_{RH}^3} 
\exp \left(-\frac{2M_X} {\kappa T_{\rm MAX}} A^{3/8} \right) 
\left( \alpha_s A^{19/8} + \alpha_p A^2\frac{\kappa T_{\rm MAX}}{M_X}\right)\ .
\label{butxxx}
\end{equation} 

We will find the solution for $X(\infty)$ as a Gaussian-integral
approximation to Eq.\ (\ref{butxxx}).  Although Eq.\ (\ref{butxxx}) is
valid only at early times, we will integrate it in the full range
between $A=0$ and $A=\infty$. This is a good approximation because the
exponential suppression makes the right-hand side of Eq.\
(\ref{butxxx}) negligible anywhere outside a small interval of scale
factors centered around $A=A_*$, with
\begin{eqnarray}
A_* & = & \left( \Frac{17}{2} \Frac{\kappa T_{\rm MAX}}{2M_X} \right)^{8/3} 
                \qquad {\rm for}\ s{\rm -wave} \ , \label{asa} \\
A_* & = & \left( \Frac{15}{2} \Frac{\kappa T_{\rm MAX}}{2M_X} \right)^{8/3} 
                \qquad {\rm for}\ p{\rm -wave} \ . \label{asb}
\end{eqnarray}
Using Eq.\ (\ref{temp}), we find that $A_*$ corresponds to a
temperature $T_*=4M_X/17$ for $s$-wave and $T_*=4M_X/15$ for
$p$-wave. Therefore $T_*$ is the temperature at which most of the
$X$-particle production takes place. We will assume $T_*<T_{\rm MAX}$
or else the final $X$-particle density is suppressed by a very small
exponential function.  The Gaussian-integral approximation to Eq.\
(\ref{butxxx}) is\footnote{The result of the Gaussian integration in
Eq.\ (\ref{xinf}) is a very good approximation of the exact integral,
which is given by $ X_\infty =(8g^2M_X^5)/(3\pi^3H_I)\left[ \alpha_s
f_s\left( 2M_X/\kappa T_{max}\right) + \alpha_p f_p\left( 2M_X/\kappa
T_{max}\right) \right]$, where $f_s(x)=\exp(-x) 8!\sum_{k=0}^8
x^{k-12}/k!  \simeq \sqrt{2\pi} \left(17/2 \right)^{17/2} \exp(-17/2)
x^{-12}$ and $f_p(x)=\exp(-x) (8!/4)\sum_{k=0}^7 x^{k-12}/k!  \simeq
2\sqrt{2\pi} \left(15/2 \right)^{15/2} \exp(-15/2) x^{-12} \simeq
f_s(x)/4$. }
\begin{equation}
X_\infty =\frac{8g^2M_X^4\sqrt{2\pi}}{3\pi^3H_IT_{RH}^3} \, 
\left( \frac{\kappa T_{\rm MAX}}{2M_X}\right)^{12}
\exp(-17/2)\left(\frac{17}{2}\right)^{17/2} 
\left(\alpha_s  +\frac{\alpha_p}{4} \right) \ ,
\label{xinf}
\end{equation}

Next we want to relate $X_\infty$ to the mass density of $X$ particles
today.  After particle production stops at $A \simeq A_*$, the factor 
$X \propto n_XA^3 = X_\infty$ remains constant. Therefore, at reheating
\begin{equation}
\rho_X (T_{RH})= M_Xn_X (T_{RH}) = M_X X_\infty T_{RH}^3 A_{RH}^{-3} \ ,
\end{equation}
where from Eq.\ (\ref{temp})
\begin{equation}
A_{RH}^{-3}=\left( \frac{T_{RH}}{\kappa T_{\rm MAX}}\right)^8
= \frac{5\pi^3g_*(T_{RH}) T_{RH}^4}{9 M_{Pl}^2 H_I^2}\ .
\end{equation}
Also, at reheating the radiation energy density is
\begin{equation}
\rho_R (T_{RH})=\frac{\pi^2}{30}g_*(T_{RH}) T_{RH}^4 \ . 
\end{equation}
After the completion of reheating the universe is radiation dominated, and 
\begin{equation}
\frac{\rho_X (T_{\rm now})}{\rho_R (T_{\rm now})}=\frac{T}{T_{\rm now}}\, 
\frac{\rho_X (T)} {\rho_R (T)}\ .
\label{nowrh}
\end{equation}
Of course the extraction of energy from the scalar field is not an
instantaneous process, but we can use $T=T_{RH}$ in Eq.\ (\ref{nowrh})
and correct for the entropy release after $T_{RH}$.  It is
straightforward to demonstrate (see the Appendix)
that relatively independent of the
model parameters, only about 25\% of the comoving $\phi$ energy
density has been extracted at $T=T_{RH}$.
At temperatures smaller than $T_{RH}$ some residual
entropy is  released by the decays of the scalar field till the
time  when the energy density in radiation  significantly dominates over the
energy density of the scalar field.
  One can show that
(again nearly independent of model parameters) the comoving entropy
increases by about a factor of 8 after $T_{RH}$.  Therefore all
the analytic estimates should be divided by a factor of 8. This
is confirmed by numerical calculations as shown in Sec.\
\ref{summary}.
The reader is referred to the Appendix for more details. 

{}From Eq.\ (\ref{nowrh}) with $T=T_{RH}$ and the extra factor of
$1/8$, we obtain an estimate for
the present energy density of $X$ particles in units of the critical
density
\begin{eqnarray}
\Omega_Xh^2 & = & 
    \Frac{3\sqrt{10}\left(17/2e\right)^{17/2} }{2048\pi^6} \, 
\Frac{g^2 g_*^{3/2}(T_{RH})}{g_*^3(T_*)} \, 
\Frac{ M_{Pl} T_{RH}^7}{M_X^7 T_{\rm now}} \, 
\left( \alpha_s+\alpha_p/4\right) \, \Omega_Rh^2 \nonumber \\
& = & 2.1\times 10^4  \left( \Frac{g}{2}\right)^2
\left[ \Frac{g_*(T_{RH})}{10}\right]^{3/2}
\left[ \Frac{10}{g_*(T_*)}\right]^3
\left( \Frac{10^3 T_{RH}}{M_X}\right)^7
\left( \alpha_s+\alpha_p/4\right)     
\label{omneq} \\ &&
\mbox{(nonrelativistic nonequilibrium production during reheating era)}
\nonumber \ . 
\end{eqnarray}
Here we have used $T_{\rm now}=2.35\times 10^{-13}$\,GeV and $\Omega_R
h^2= 4.17 \times 10^{-5}$, including the contributions from the cosmic
background radiation and from neutrinos.  

Notice that in this case
$\Omega_X$ is proportional to the annihilation cross section, instead
of being inversely proportional, as in the case of the usual thermal
relic abundance calculation in a radiation-dominated universe.

The basic assumption used in this section is that the $X$ particles
never reach thermal equilibrium.  This hypothesis holds if at the time
of maximum particle production the $X$ number density is less than the
equilibrium value, $X_\infty <X_{eq}(T_*)$. Using Eqs.\ (\ref{eqnox})
and (\ref{xinf}), we find that this condition corresponds to a limit
on the annihilation cross section $\alpha_s<{\bar \alpha_s}$ for
$s$-wave or $\alpha_p<{\bar \alpha_p}$ for $p$-wave, with
\begin{eqnarray}
\bar \alpha_s &=& \frac{4\sqrt{10} \pi^{5/2} e^{17/4} }{289} \, 
\frac{ g_*(T_*) }{g g_*^{1/2}(T_{RH})} \,
\frac{M_X^3}{M_{Pl} T_{RH}^2}
\nonumber \\
& = &  7\times 10^{-10} \, \frac{2}{g} \left[ \frac{g_*(T_*)}{10}\right]
\left[ \frac{10}{g_*(T_{RH})}\right]^{1/2}
\left( \frac{M_X}{\rm 100\, GeV}\right)^3
\left( \frac{{\rm 100\, MeV}}{T_{RH}}\right)^2  , \label{pirl1}\\
\bar \alpha_p &=& \frac{2\sqrt{10} \pi^{5/2} e^{15/4} }{15}
\frac{ g_*(T_*)}{g g_*^{1/2}(T_{RH})} \,
\frac{M_X^3}{M_{Pl} T_{RH}^2} \nonumber \\
&=&  4\times 10^{-9} \, \frac{2}{g}
\left[ \frac{g_*(T_*)}{10}\right]
\left[ \frac{10}{g_*(T_{RH})}\right]^{1/2}
\left( \frac{M_X}{\rm 100\,GeV}\right)^3
\left( \frac{{\rm 100\,MeV}}{T_{RH}}\right)^2  \label{pirl2} \ .
\end{eqnarray}

In the non-equilibrium case considered in this section, we find that
there is a maximum value of $\Omega_X$ that can be achieved. This is
obtained by replacing the constraint $\alpha_s<\bar{\alpha}_s$ into Eq.\
(\ref{omneq}),
\begin{equation}
\Omega_Xh^2 < 1\times10^{-5} \, \frac{g}{2} \left[ 
\frac{g_*(T_{RH})}{10}\right]
\left[ \frac{10}{g_*(T_*)}\right]^2
\left( \frac{T_{RH}}{\rm 100\,MeV}\right)^5
\left( \frac{\rm 100\,GeV}{M_X}\right)^4  \ .
\label{lim1}
\end{equation}
A similar constraint can be obtained in the case of $p$-wave
annihilation.

To conclude this section, we want to show that for the case under
consideration the process of particle production always freezes out
before reheating, {\it i.e.,}
\begin{equation}
H(A_*)>\Gamma_\phi \ .
\label{hgam}
\end{equation}
Since the universe is matter dominated by the $\phi$ field, $H$ scales
like $A^{-3/2}$ and therefore $H(A_*)=H_IA_*^{-3/2}$. Using Eqs.\
(\ref{asa})--(\ref{asb}), we find that the relation in Eq.\
(\ref{hgam}) is satisfied whenever $M_X>3T_{RH}$. This condition is
always verified once we assume that the $X$ particles do not
thermalize after reheating.

\subsection{Nonrelativistic Equilibrium Production and Freeze Out} \label{nreq}

We will now consider the case in which the annihilation cross section
is large ($\alpha_s>\bar{\alpha}_s$ or $\alpha_p>\bar{\alpha}_p$) and
the $X$ particle species reaches chemical equilibrium before reheating
(this case is also discussed in Ref.\ \cite{jmd}).  The calculation of
$\Omega_X$ is now analogous to the ordinary calculation of thermal
relic abundances.  However, the result is different because the
relation between temperature and scale factor is not the same as in
the ordinary case of a radiation-dominated universe.

The freeze-out temperature $T_F$ is obtained by solving for the condition
\begin{equation}
n_X^{eq} (T_F)\langle \sigma v\rangle = H(T_F)  \ .
\label{eqcaz}
\end{equation}
Here $n_X^{eq}=X_{eq}A^{-3}T_{RH}^3$ is the equilibrium number density
of $X$ particles and the expansion rate $H$ as a function of
temperature was expressed in Eq.\ (\ref{hhhh}).

Defining $x_F\equiv M_X/T_F$, the condition in Eq.\ (\ref{eqcaz}) can
be written as
\begin{equation}
x_F=\ln \left[ \frac{3}{2\sqrt{10}\pi^3} \, 
\frac{g g_*^{1/2}(T_{RH})}{g_*(T_F)} \,
\frac{M_{Pl}  T_{RH}^2}{M_X^3} \,
(\alpha_s x_F^{5/2} + 5 \alpha_p x_F^{3/2} /4) \right] \ .
\label{solut}
\end{equation}
The factor $5/4$ in front of the $p$-wave term has been added to match
the analytic solution of the Boltzmann equation.  Notice that Eq.\
(\ref{solut}) admits a solution only for $\alpha_s>1.5 \bar \alpha_s$
or $\alpha_p >0.4 \bar \alpha_p$, in nice agreement with the starting
assumption on the annihilation cross section.

For comparison, we remind that in the case of a radiation-dominated
universe the expansion rate is
$H=\sqrt{4\pi^3 g_*/45} \, T^2/M_{Pl}$ and the analog to Eq.\ (\ref{eqcaz}) is
\begin{equation}
x_F=\ln \left[ \frac{3\sqrt{5}}{4\sqrt{2}\pi^3} \, 
\frac{g}{g_*^{1/2}(T_F)} \, \frac{M_{Pl}}{ M_X} \,
(\alpha_s x_F^{1/2} + 2\alpha_p x_F^{-1/2})
\right] \ .
\label{solus}
\end{equation}

In order to compute $\Omega_X$ we can use Eq.\ (\ref{nowrh}) (again
with the understanding that there will be an overall correction due to
the fact that the reheating process is not complete at $T_{RH}$) to
derive the result
\begin{equation}
\rho_X(T_{RH})=\left(\frac{g_*(T_{RH})}{g_*(T_{F })}\right)^2                 
\left( \frac{T_{RH}}{T_F}\right)^8 \rho_X^{eq}(T_F)
=\frac{g}{(2\pi)^{3/2}} \,\left(\frac{g_*(T_{RH})}{g_*(T_{F })}\right)^2  
\,  \frac{T_{RH}^8}{M_X^4}  \,  x_F^{13/2}  
\exp(-x_F) \ .
\end{equation}
Thus we obtain, in the case decoupling occurs during reheating, the result
\begin{eqnarray}
\Omega_X h^2 &=& \frac{5\sqrt{5}}{4\sqrt{\pi}}  \,  
\frac{g^{1/2}_*(T_{RH})}{g_*(T_{F })}  \,
\frac{T_{RH}^3}{T_{\rm now} M_X  M_{Pl} } \,
\frac{1}{(\alpha_s x_F^{-4} + 4\alpha_p x_F^{-5}/5)} \, \Omega_R h^2
\nonumber \\
&=& 2.3\times 10^{-11} \, \frac{g^{1/2}_*(T_{RH})}{g_*(T_{F })}  \, 
\frac{T_{RH}^3  {\rm GeV}^{-2}}
{  M_X  (\alpha_s x_F^{-4} +4\alpha_p x_F^{-5}/5)} 
\label{oom}\\
& & \mbox{(nonrelativistic equilibrium production during reheat era)} 
\nonumber \ .
\end{eqnarray}
In this case $\Omega_X$ is inversely proportional to the annihilation
cross section, as in the radiation-dominated case.  Eq.\ (\ref{oom})
generalizes the ordinary result of decoupling during the
radiation-dominated era, which is given by
\begin{eqnarray}
\Omega_X h^2 &=&\frac{4\sqrt{5} \, \Omega_R h^2  \, M_X^2}{\sqrt{\pi}  \,
T_{\rm now}  g_*^{1/2}(T_{F}) M_{Pl}  (\alpha_s x_F^{-1} +\alpha_p x_F^{-2}/2)}
\nonumber \\
&=& 7.3\times 10^{-11} \, \frac{1}{g_*^{1/2}(T_{F})} \, 
\frac{{\rm GeV}^{-2}}{M_X^{-2}(\alpha_s x_F^{-1} +\alpha_p x_F^{-2}/2)} 
\label{oomc} \\
& & \mbox{(nonrelativistic production during radiation era)}
\nonumber \ . 
\end{eqnarray}
Indeed, Eq.\ (\ref{oom}) approximately
reduces to Eq.\ (\ref{oomc}) as $T_{RH}$
approaches $T_F$. Furthermore, Eq.\ (\ref{oom}) also reproduces Eq.\
(\ref{omneq}) when $\alpha_s$ approaches ${\bar \alpha_s}$ from above.

The effect of a low reheat temperature is to reduce the relic
abundance with respect to the ordinary case by a factor
$T_{RH}^3T_F^{\rm old}/ (T_F^{\rm new})^4$, where $T_F^{\rm old}$ and
$T_F^{\rm new}$ are the freeze-out temperatures in cosmologies with
high and low $T_{RH}$, respectively.  This suppression factor can be
understood in the following way. During the epoch before reheating,
the expansion is faster than in the radiation-dominated era;
freeze out occurs earlier, enhancing the $X$ abundance at $T=T_F$ [see
Eqs.\ (\ref{solut}) and (\ref{solus})].  However, as the universe cools
from $T_F$ to $T_{RH}$, the expansion dilutes $n_X$ by a factor
$(T_{RH}/T_F)^8$; the dilution is more effective than in the
matter-dominated case ($n_X \sim T^3$) because of entropy release
during reheating.  This explains why $\Omega_X$ is roughly
$T_{RH}^3T_F^{\rm old}/ (T_F^{\rm new})^4$ times the relic density
obtained in the case of large reheat temperature.

\begin{figure}[t]
\centering \leavevmode\epsfxsize=350pt \epsfbox{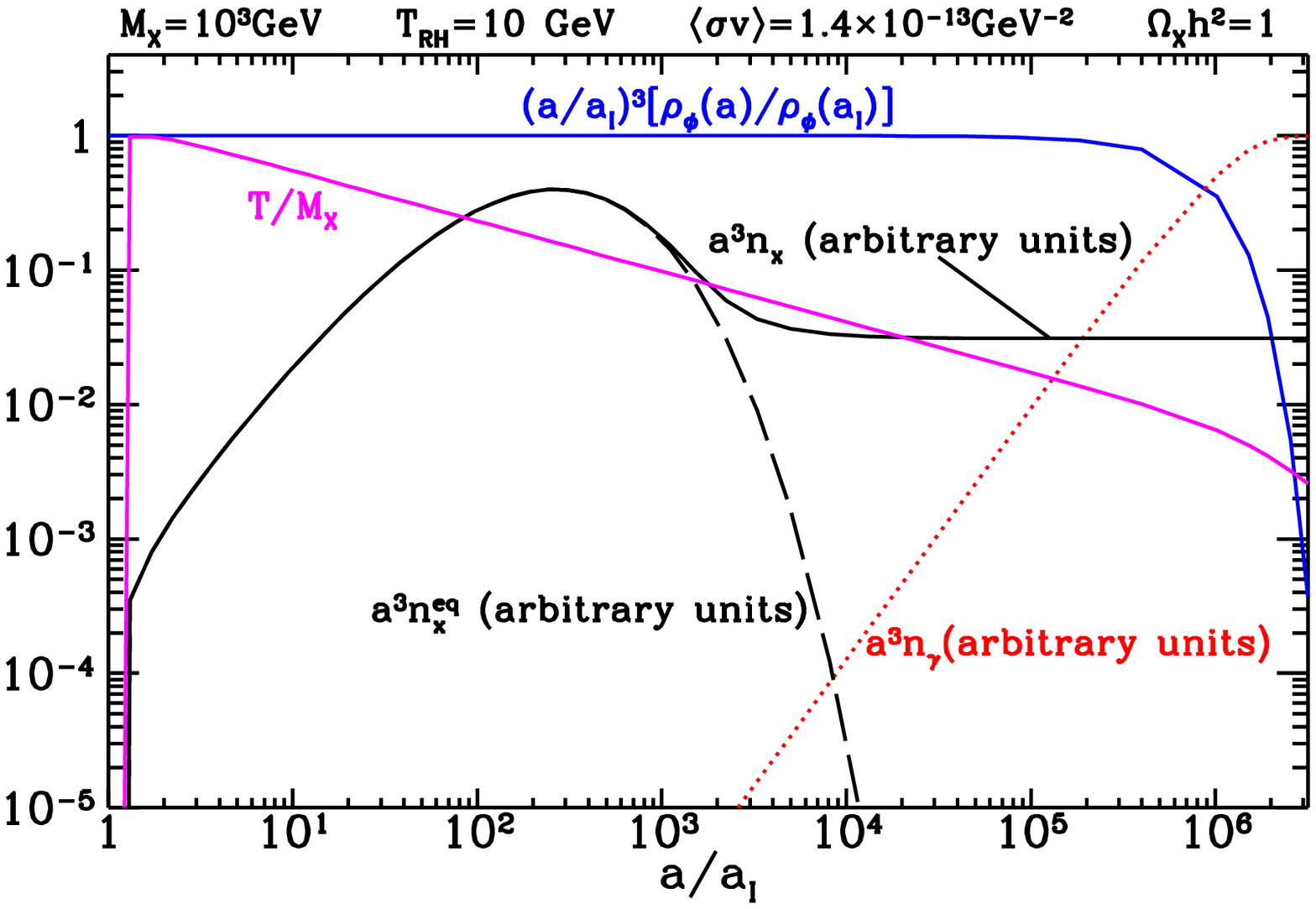}\\
\centering \leavevmode\epsfxsize=350pt \epsfbox{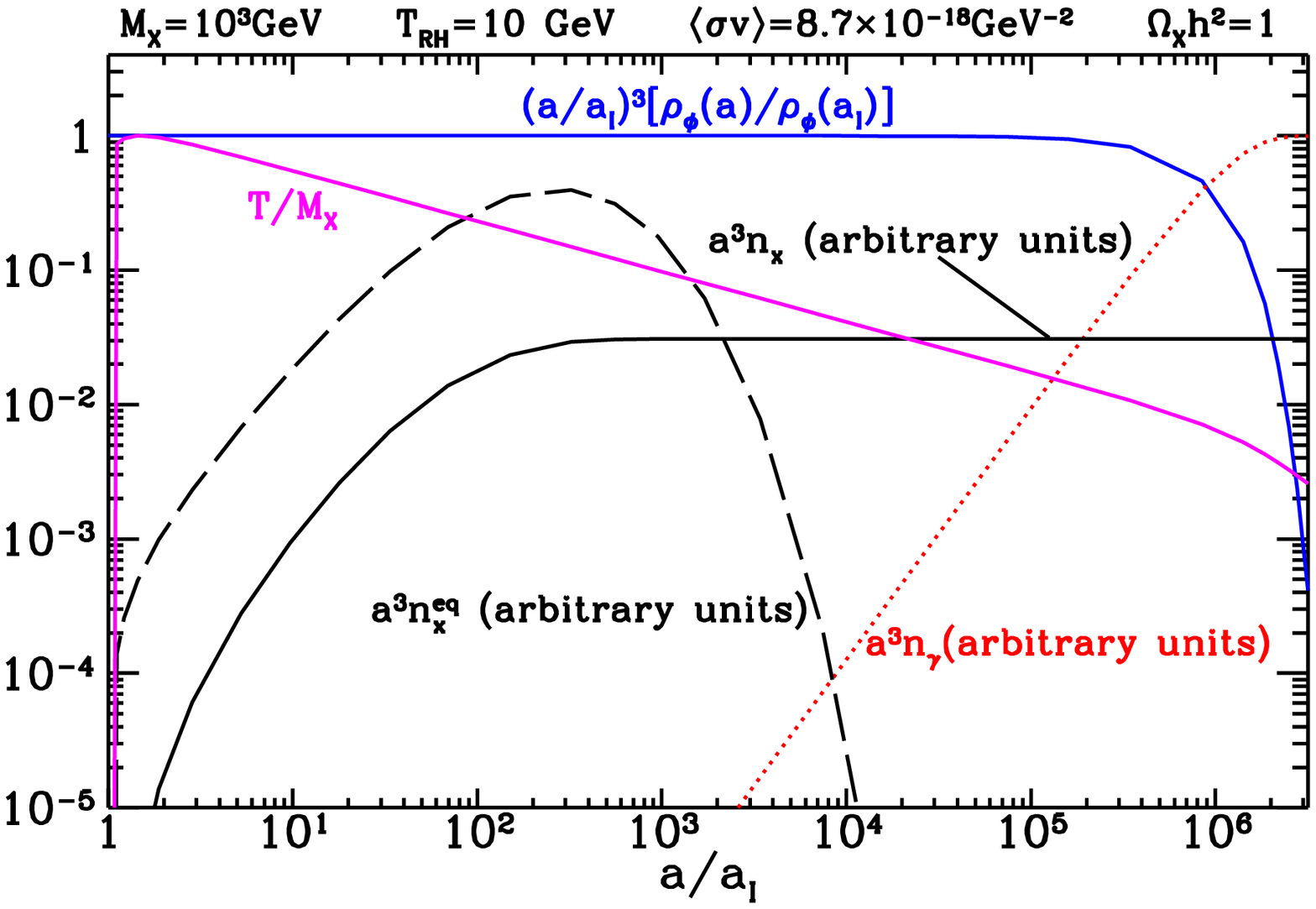}
\caption{Shown in the upper graph is the evolution of the $X$ density 
in the case that the cross section is sufficiently large to establish
chemical equilibrium prior to freeze out. The lower graph illustrates
the case where the cross section is too small to establish chemical
equilibrium. The two cross sections were chosen to result in the same
final $X$ abundances necessary to give a critical density of $X$ particles
today. In the calculations $g_*$ was kept constant at $g_*=30$.\label{eqnoneq}}
\end{figure}

An illustration of the freeze out of the $X$-abundance in equilibrium
and out of equilibrium is illustrated in Fig.~\ref{eqnoneq}.  In both
cases the final $X$-abundance is the same.  In the top graph the cross
section is large enough to establish equilibrium prior to freeze out,
while in lower graph the cross section is too small to establish
equilibrium.

\subsection{Nonrelativistic Production and Freeze Out}  
\label{nonrel}

The relic density calculations performed in Secs.\ \ref{nrnoneq} and
\ref{nreq}, under the assumptions of out-of-equilibrium and
equilibrium respectively, match well in the intermediate region in
spite of the fact that they are derived with different approximations.
Indeed, if we use in Eq.\ (\ref{oom}) the minimum allowed cross
section ($\alpha_s=\bar \alpha_s$), corresponding to the maximum
allowed freeze-out temperature ($x_F=5/2$), we obtain the bound
\begin{equation}
\Omega_Xh^2 < 4\times 10^{-6} \, \frac{g}{2} 
\left[ \frac{10}{g_*(T_{RH})}\right]
\left( \frac{T_{RH}}{\rm 100\,MeV}\right)^5
\left( \frac{\rm 100\,GeV}{M_X}\right)^4 .
\label{lim2}
\end{equation}
When this bound is saturated, we are approaching the transition from
the results of Sec.\ \ref{nreq} to those of Sec.\ \ref{nrnoneq}. Indeed,
the results in Eqs.\ (\ref{lim1}) and (\ref{lim2}) turn out to be in fair
agreement with each other.  Similar conclusions can be obtained in the
case of $p$-wave annihilation, but for simplicity in this section we
will consider only the case of dominant $s$-wave annihilation.

\begin{figure}[t]
\centering \leavevmode\epsfxsize=350pt \epsfbox{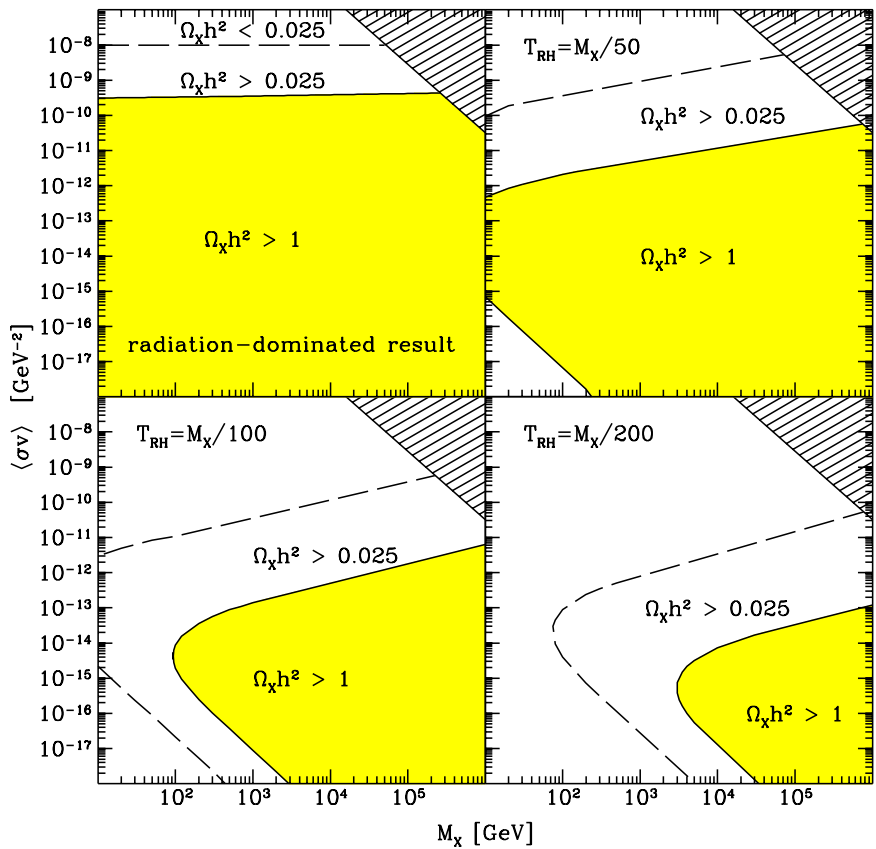}
\caption{The shaded areas show the cosmologically excluded regions
for a particle of mass $M_X$ with 2 degrees of freedom which
annihilates in $s$-wave with a thermal-averaged nonrelativistic cross
section $\langle \sigma v \rangle$. The upper-left figure is the usual
case where particle freeze out occurs when the universe is radiation
dominated. In the other frames, we have chosen 
$M_X/T_{RH}=50$, 100, and 200. The interesting region for cold dark
matter ($0.025<\Omega_X h^2<1$) is between the dashed line and the
shaded area.  The upper right-hand corner of the $M_X- \langle \sigma
v \rangle$ plane is excluded by unitarity arguments.  \label{exclude}} 
\end{figure}

In the ordinary case of large reheat temperature ($T_{RH}\simgt
M_X$, {\it i.e.,} production and freeze out in a radiation-dominated
universe), $\Omega_Xh^2$ is proportional to $\langle \sigma
v\rangle^{-1}$, as can be seen from Eq.\ (\ref{oomc}).  So except for
a logarithmic correction, the is no explicit mass dependence to
$\Omega_X h^2$.  The constraint from the age of the universe,
$\Omega_X h^2 \simlt 1$, implies a lower bound on the $X$ annihilation
cross section, as shown in the upper-left graph in Fig.\  \ref{exclude}. The
unitarity limit to $\langle \sigma v\rangle$ as a function of the
mass, $\langle \sigma v\rangle_{\rm MAX} = 8\pi/M_X^2$, is also
shown. The two bounds cross at the value $M_X =340$\,TeV \cite{kam},
which is usually taken as a cosmological upper bound on any stable
massive particle.

In Fig.\ \ref{exclude} we show how the cosmological bounds on $M_X$
can be relaxed for sufficiently low $T_{RH}$. As $T_{RH}$ is
decreased, the allowed region in the parameter space $\langle \sigma
v\rangle$ versus $M_X$ grows.  The numerical results presented in the
figure (assuming $g_*=30$) are in excellent agreement with the
analytic estimates.  The results are simple to understand.

In the case of low reheat temperature, the constraint $\Omega_X
h^2<1$ does not simply give a lower bound on $\langle \sigma
v\rangle$, but rules out a range of values. Indeed, for very small
$\langle \sigma v\rangle$ we are in the limit of nonrelativistic
nonequilibrium production discussed in Sec.\ \ref{nrnoneq}, and Eq.\
(\ref{omneq}) applies.  For fixed $T_{RH}/M_X$, $\Omega_Xh^2 \propto
M_X^2\langle \sigma v \rangle$, and the value of $\langle \sigma v
\rangle$ forming contours of constant $\Omega_Xh^2$ will scale as
$M_X^{-2}$.  As $\langle\sigma v\rangle$ grows for fixed $M_X$,
$\Omega_X$ increases and eventually may conflict with the age of the
universe constraint.  After the maximum value of $\Omega_X$ is
reached, see Eq.\ (\ref{lim1}), a further increase of $\langle \sigma
v\rangle$ brings us to the limit of nonrelativistic equilibrium
production of Sec.\ \ref{nreq}. Now $\Omega_X$ is obtained from Eq.\
(\ref{oom}), and the value of $\langle \sigma v\rangle$ to give
contours of fixed $\Omega_Xh^2$ increases (albeit slowly) as $M_X$
grows.

Figure \ref{exclude} also shows the parameter region in which the
particle $X$ could be an interesting cold dark matter candidate,
$0.025<\Omega_X h^2<1$. Values of annihilation cross sections and
masses which are ordinarily excluded in the case of large $T_{RH}$ can
now be of particular cosmological and observational interest.

\begin{figure}[t]
\centering \leavevmode\epsfxsize=350pt \epsfbox{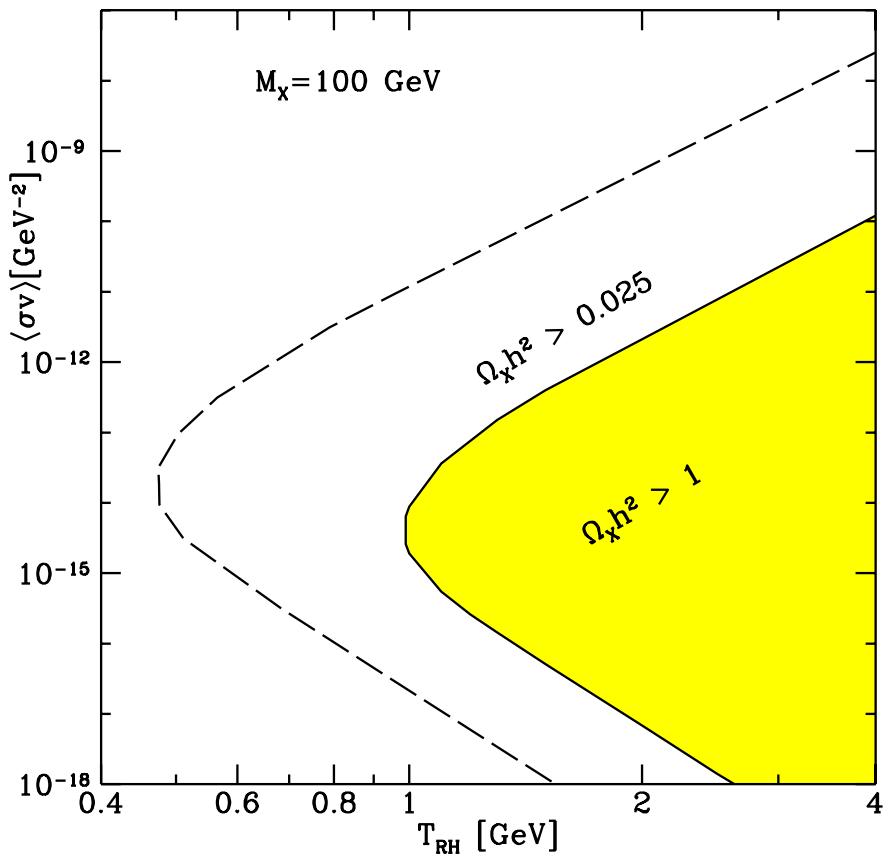}
\caption{The shaded region shows the cosmologically excluded region,
as a function of the reheat temperature $T_{RH}$, for a particle of
mass $M_X=100$\,GeV with 2 degrees of freedom which annihilates with a
thermal-averaged nonrelativistic $s$-wave cross section $\langle
\sigma v \rangle$. The region interesting for cold dark matter
($0.025<\Omega_X h^2<1$) is delimited between the dashed line and the
shaded area.
\label{figreheat}} 
\end{figure}

In Fig.\ \ref{figreheat} we show the cosmologically excluded region
($\Omega_X h^2>1$) and the region relevant for dark matter
($0.025<\Omega_X h^2<1$) as a function of $T_{RH}$ for a fixed value
$M_X=100$\,GeV. Notice how the lower limit on $\langle \sigma
v\rangle$, which is $2\times 10^{-10}$\,GeV$^{-2}$ in the ordinary
radiation-dominated cosmology, is relaxed as $T_{RH}$ is lowered.

\begin{figure}[t]
\centering \leavevmode\epsfxsize=350pt \epsfbox{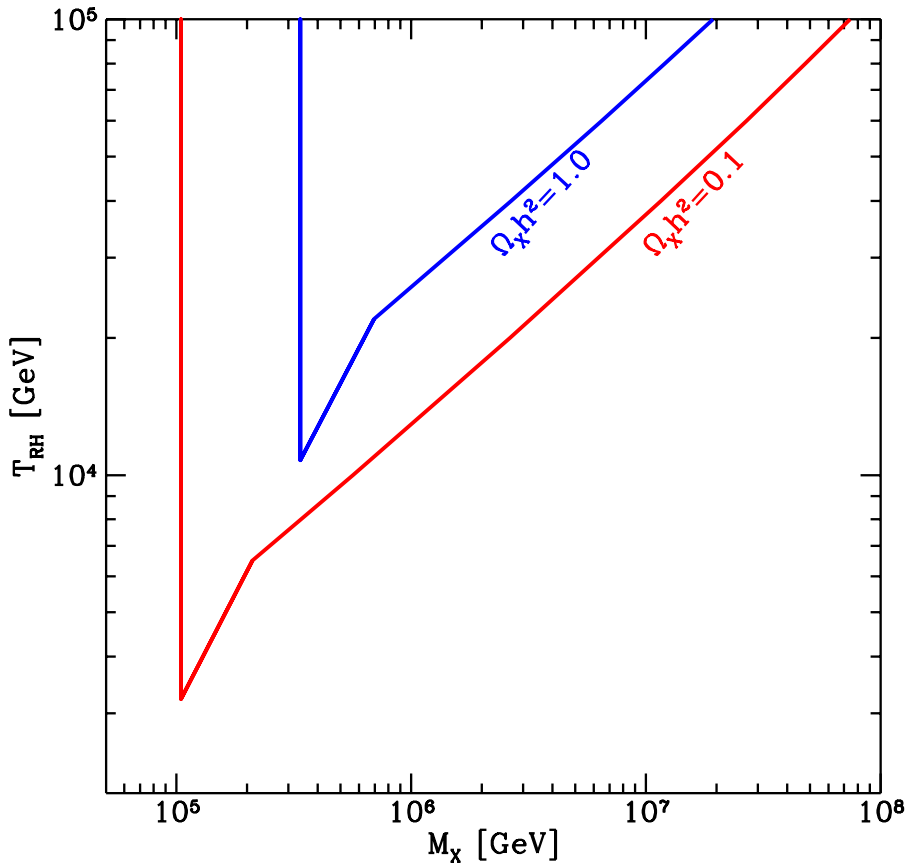}
\caption{The relic abundance $\Omega_X h^2$ as a function of the
reheat temperature $T_{RH}$ for a particle with 2 degrees of
freedom, mass $M_X$ and a nonrelativistic annihilation cross section
in $s$-wave saturating the unitarity bound. 
\label{figunitarity}}
\end{figure}

A striking implication of dark matter production during reheating is
that the unitarity bound, $M_X <340$\,TeV, disappears and one could
conceive thermal relics of very heavy particles without conflicting
with the age of the universe. The necessary assumption is that $T_{\rm
MAX}$ is larger than the temperature at which the relevant physical
processes occur, {\it i.e.,} $T_*$ in the out-of-equilibrium case and
$T_F$ in the equilibrium case. Once this assumption is made, the final
result on $\Omega_X$ does not depend $T_{\rm MAX}$ or other any
initial conditions of the inflationary model, but only on
$T_{RH}$. Figure \ref{figunitarity} shows how the unitarity bound is
modified in presence of low $T_{RH}$. Here we have taken
$\alpha_s=8\pi$ and plotted contours of various values of
$\Omega_Xh^2$ as a function of $M_X$ and $T_{RH}$.

\subsection{Relativistic $X$}  \label{rel}

In this section we consider the case in which the relevant physical
processes of particle production and freeze out occur when $X$ is
still relativistic, and therefore we assume that the inequality of Eq.\
(\ref{ineq}) is not satisfied. If the annihilation cross section is
not suppressed by any mass scale larger than $M_X$, then $\langle
\sigma v \rangle \sim T^{-2}$ and $X$ remains in thermal equilibrium
until it becomes nonrelativistic. More interesting is the case in
which the annihilation process depends on a new mass scale $M_G \gg
M_X$. Therefore, we define
\begin{equation}
\langle \sigma v \rangle \equiv \frac{T^n}{M_G^{n+2}},
\end{equation}
for a generic exponent $n$.  As concrete examples, one can think of a
heavy neutrino or a neutralino in the regime in which the temperature
is larger than their masses: in this case $n=2$ and $M_G$ is roughly
the mass of an intermediate gauge boson or of a slepton,
respectively. Another interesting example is the gravitino, for which
$n=0$ and $M_G\sim M_{Pl}$. Finally, one can consider the graviton
Kaluza-Klein excitations in theories with $\delta$ large extra
dimensions \cite{qg}, for which $n=\delta$ and $M_G$ is of the order
of the fundamental gravitational scale.

Let us first consider the case in which $X$ does not reach an
equilibrium density. The analysis is similar to the one performed in
Sec.\ \ref{nrnoneq}. At early times, the Boltzmann equation can be
approximated by
\begin{equation}
\frac{dX}{dA}= \sqrt{\frac{3}{8\pi}} A^{-5/2} 
\langle \sigma v\rangle M_{Pl} T_{RH} \Phi_I^{-1/2} X_{eq}^2 \ , \qquad
\label{pirl}
X_{eq}=\frac{c_\xi}{\pi^2} \left(\frac{T}{T_{RH}}\right)^3 A^3 \ .
\end{equation}
where $c_\xi=g \xi(3)$ for bosons, $c_\xi=(3/4)g \xi(3)$ for fermions,
and $g$ is the number of $X$ degrees of freedom.  Using the relation
between scale factor and temperature in Eq.\ (\ref{temp}), we can
rewrite Eq.\ (\ref{pirl}) as
\begin{equation}
\frac{dX}{dT}= - \frac{8}{\sqrt{5\pi^{11}}} \, \frac{g_*^{1/2}(T_{RH})}{g_*(T)}
               \, c_\xi^2  
               \, \frac{(\kappa T_{max})^8M_{Pl}}{T^{7-n}M_G^{2+n}T_{RH}} \ .
\label{fric}
\end{equation}
For $n<6$ the $X$-particle production dominantly occurs at the lowest
possible temperature. In all interesting situations we know of, the
annihilation cross section is such that $n<6$, and therefore we
consider only this case. Integrating Eq.\ (\ref{fric}) up to a final
temperature $T_f$, we obtain for $n<6$,
\begin{equation}
n_X(T_f)= \frac{1}{6-n} \,
          \frac{8}{\sqrt{5\pi^{11}}} \, \frac{g_*^{1/2}(T_{RH})}{g_*(T_f)} \,
           c_\xi^2 \, \left(\frac{T_f}{M_G}\right)^{n+2} \, M_{Pl}T_{RH}^2 \ .
\label{risn}
\end{equation}

If $M_X>T_{RH}$, Eq.\ (\ref{risn}) should be evaluated at $T_f=M_X$
and the result be used as an initial condition for the
nonrelativistic analysis. Since we have linearized the differential
equation, within our approximation this simply amounts to adding the
relic density obtained from Eq.\ (\ref{risn}) to the contribution
derived in Sec.\ \ref{nrnoneq}.

If $M_X<T_{RH}$, as it is usually the case for the gravitino, then in
Eq.\ (\ref{risn}) we can take $T_f=T_{RH}$.  At temperatures smaller
than $T_{RH}$, the universe is radiation dominated and a calculation
analogous to the one that led us to Eq.\ (\ref{risn}) shows that in
this case $X$-particle production dominantly occurs at the largest
possible temperature, as long as $n>-1$.  Therefore all the relevant
dynamics occurs at $T=T_{RH}$. The relic abundance can be obtained by
rescaling the $X$ number density in Eq.\ (\ref{risn}) to the present
temperature, as done in Sec.\ \ref{nrnoneq}, to yield
\begin{eqnarray}
\Omega_X h^2 & = & \frac{48\sqrt{5}}{(6-n)\pi^{15/2}}  \,
\frac{c_\xi^2}{g_*^{3/2}(T_{RH})} \, \frac{ M_{Pl} M_X T_{RH}^{n+1}}
{  T_{\rm now} M_G^{n+2}} \,  \Omega_R h^2 \nonumber \\
&=& \left[ \frac{c_\xi}{3\xi(3)/2}\right]^2 \left[ \frac{10}{g_*(T_{RH})}
\right]^{3/2}  \frac{M_X T_{RH}^{n+1}}{M_G^{n+2}} \, 
\frac{4\times 10^{24}}{(6-n)} \ .
\label{opsm}
\end{eqnarray}

The result in Eq.\ (\ref{opsm}) is valid as long as the $X$ particles
do not reach equilibrium at temperatures larger than $T_{RH}$, or
$n_X(T_{RH})<n_X^{eq}(T_{RH})$. This implies
\begin{equation}
M_G> \left[ \frac{8 c_\xi M_{Pl} T_{RH}^{n+1}}{
\sqrt{5} (6-n) \pi^{7/2} g_*^{1/2}(T_{RH})} \right]^{1/n+2} .
\label{piffer}
\end{equation}
This condition holds for both gravitinos and Kaluza-Klein gravitons.
For a Majorana fermion with $n=2$, it requires
\begin{equation}
M_G> \left[ \frac{10}{g_*(T_{RH})}\right]^{1/8} \left( \frac{T_{RH}}{\rm GeV}
\right)^{3/4}  18\,{\rm TeV} \ .
\end{equation}

If the condition (\ref{piffer}) is not satisfied, then the $X$
particle density thermalizes. However, in this case, they do not
freeze out before reheating. Indeed, let us consider the ratio of the
interaction rate versus the expansion rate at the temperature $T$
\begin{equation}
\frac{n_X^{eq}\langle \sigma v \rangle}{H} =
\frac{3c_\xi   g_*^{1/2}(T_{RH})}{\sqrt{5} \pi^{7/2} g_*(T) } \, 
\frac{M_{Pl} T_{RH}^2 T^{n-1}}{M_G^{n+2}} \ .
\end{equation}
By requiring that the condition (\ref{piffer}) does not hold and that
$T>T_{RH}$, we find
\begin{equation}
\frac{n_X^{eq}\langle \sigma v \rangle}{H} >\frac{3(6-n)}{8} \ .
\label{spuf}
\end{equation}
Since for $n<6$ the right-hand side is of order unity, Eq.\
(\ref{spuf}) shows that if a relativistic $X$ reaches thermal
equilibrium, then it does not freeze out before reheating.

In conclusion, for relativistic particles the relevant processes
determining their relic abundances occur at reheating or afterwards.
This, in particular, is true for gravitinos and for the graviton
Kaluza-Klein excitations with $\delta<6$ for which the cosmological
bound derived in Ref.\ \cite{Arkani-Hamed:1999nn} applies.

\subsection{Summary of the Different Cases}  \label{summary}

Because of all the various cases encountered, it is probably useful to
summarize the different possibilities. Let us consider a stable
weakly-interacting particle $X$ with mass $M_X$ and dominant $s$-wave
annihilation.

When $M_X \simgt 17T_{\rm MAX}/4 $ we are in the deep
nonrelativistic regime, and the $X$ relic abundance is strongly
suppressed by an exponential factor.

\begin{figure}[t]
\centering \leavevmode\epsfxsize=350pt \epsfbox{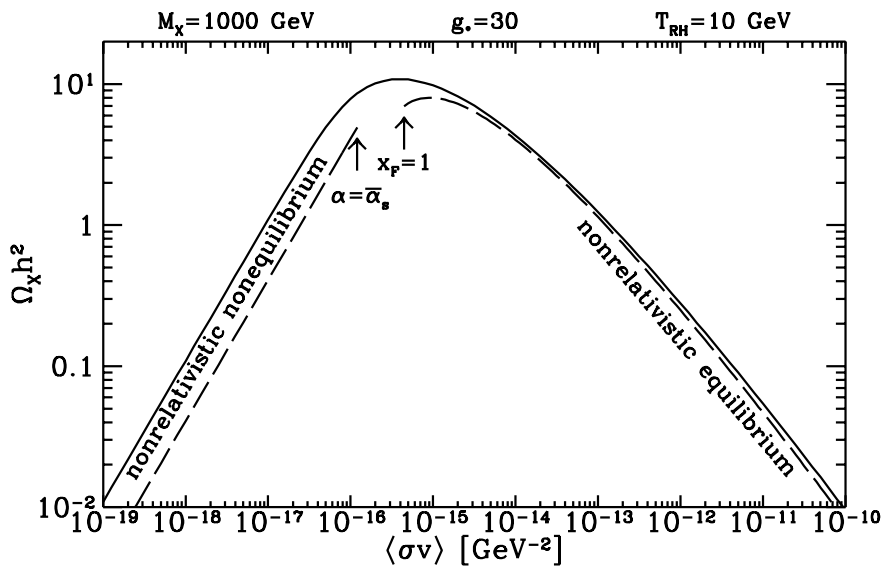}
\caption{Comparison of numerical vs.\ analytic results. The
nonequilibrium calculation is relevant for $\alpha<\bar{\alpha}$,
shown in the figure. The equilibrium calculation assumes decoupling
while nonrelativistic, or $x_F>1$.  \label{compare}}
\end{figure}

For $17T_{\rm MAX}/4 \simgt M_X \simgt T_{\rm MAX}$ we are in the
nonrelativistic case. Depending on the value of the annihilation cross
section, $X$ may or may not reach an equilibrium distribution before
freezing out. In the first case ($\alpha_s>\bar{\alpha}_s$),
$\Omega_X$ is given by Eq.\ (\ref{oom}), and in the second one
($\alpha_s<\bar{\alpha}_s$), it is given by Eq.\ (\ref{omneq}).  The
agreement between analytic estimates and numerical integration of the
Boltzmann equations is illustrated in Fig.\ \ref{compare}.

For $T_{\rm MAX} \simgt M_X \simgt x_F T_{RH}$, where $x_F$ is given in 
Eq.\ (\ref{solut}), the particle $X$ is first relativistic, then
becomes nonrelativistic and finally decouples before reheating.
The relic abundance is again given by Eq.\ (\ref{oom}).

Lighter $X$ particles thermalize after reheating, erasing any previous
information on their number density. The relic abundance has the
ordinary expression given in Eq.\ (\ref{oomc}).

We have also discussed in Sec.\ \ref{rel} the case of very light
particles with annihilation cross section suppressed by a heavy mass
scale.  In all cases of interest, the relic abundance is determined by
the physics at $T_{RH}$.

\section{Applications to Dark Matter Candidates}  \label{applications}

\subsection{Supersymmetry}  \label{susysect}

The neutralino is the most natural cold dark matter candidate in the
context of supersymmetric extensions of the Standard Model. If the
neutralino is dominantly a Higgsino its relic density is typically
small, because of the efficient coannihilation with other neutralinos
and charginos which turn out to be almost degenerate in mass.
Upcoming LEP2 runs will be able to probe the small window left
unexplored in which a light Higgsino could give a significant
contribution to the present energy density of the universe
\cite{ellisganis}.  Moreover, if the Higgsino is heavier than the
gauge bosons, the annihilation channels into $W^\pm$ and $Z^0$
strongly deplete its relic abundance. Significant contributions to
$\Omega$ then require a lightest supersymmetric particle (LSP) heavier than
about 500 GeV,
weakening the motivations for low-energy
supersymmetry.

The case of a mainly $B$-ino lightest neutralino is much more
promising for dark matter. First of all, we should recall that most of
the supersymmetric models obtained from supergravity usually predict
that the Higgs mixing parameter $\mu$ is large. This is because $\mu$
should compensate the large radiative corrections to the Higgs mass
parameters in order to achieve the correct size of electroweak
symmetry breaking.  Therefore, a common expectation is that the lightest
neutralino is an almost pure $B$-ino.

In the early universe, the $B$-ino will mainly annihilate into fermion
pairs through $t$-channel exchange of squarks and sleptons.
Exceptions occur only for pathological situations in which there is a
resonant $s$-channel exchange of $Z^0$ or a Higgs boson. Actually,
because of the large hypercharge of the right-handed electron and the
expected lightness of sleptons compared to squarks, it is often a good
approximation to include in the annihilation cross section only the
exchange of the right-handed sleptons. Summing over three slepton
degenerate families with mass ${\tilde m}_{\ell_R}$, the $B$-ino
annihilation cross section parameters are
\begin{equation}
\alpha_s=0, \quad \alpha_p=\frac{24\pi\alpha^2}{\cos^4\theta_W}
\left( 1+\frac{{\tilde m}_{\ell_R}^2}{M_B^2} \right)^{-2}.
\end{equation}
Notice that the annihilation process is $p$-wave suppressed because
of the Majorana nature of the neutralino.

Cosmological considerations give an upper bound to the $B$-ino
mass. Indeed, the requirement that charged particles are not the LSP
implies ${\tilde m}_{\ell_R}>M_B$. The minimum allowed $B$-ino relic
abundance corresponds to the maximum annihilation cross section and
therefore to the minimum ${\tilde m}_{\ell_R}$. Setting ${\tilde
m}_{\ell_R} =M_B$ in the expression for $\Omega$, one obtains an upper
bound on the $B$-ino mass of about 300 GeV (for $\Omega h^2<0.3$)
\cite{bees}.  The bound can be weakened in the presence of 
resonant $s$-channel annihilations, once a small Higgsino admixture
is introduced. Moreover, in constrained models in which the
supersymmetry-breaking masses satisfy simple universal relations at
the GUT scale, this bound reduces to 200 GeV \cite{kane}.

However, as emphasized in Ref.\ \cite{falk}, whenever the sleptons and
the $B$-ino become degenerate in mass within about 10--20{\%}, one
cannot ignore the effects of coannihilation. These effects can modify
significantly the $B$-ino relic abundance, because annihilation
channels involving the charged sleptons have large cross sections
which are not $p$-wave suppressed. Indeed, even in the case of the
constrained model, the previous limit on the $B$-ino mass can be
relaxed to about 600 GeV \cite{falk}.

\begin{figure}[t]
\centering \leavevmode\epsfxsize=350pt \epsfbox{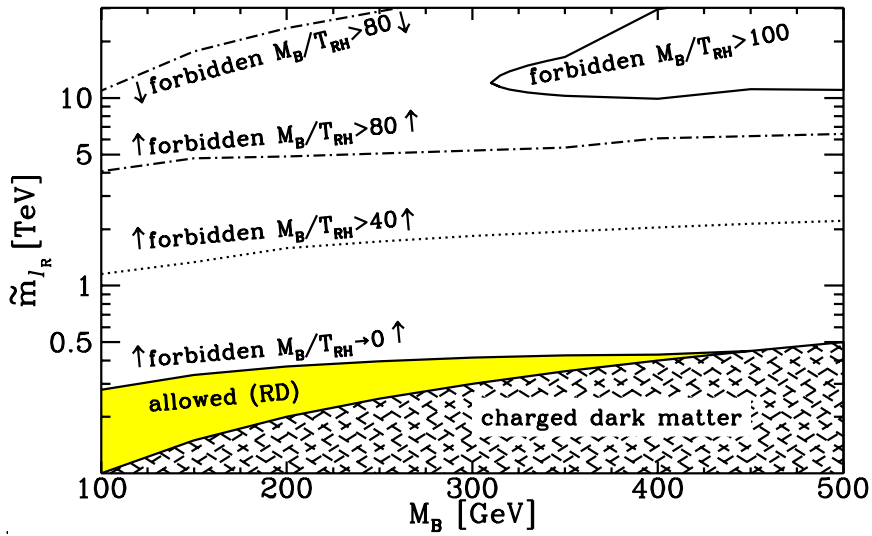}
\caption{The allowed region in the $M_B$--${\tilde m}_{\ell_R}$ plane 
in the standard cosmology where freeze out occurs in the
radiation-dominated era is indicated by the shaded region above the
disallowed region where the LSP is charged and below the curve marked
``forbidden $M_B/T_{RH}\rightarrow 0$'' above which $\Omega_Bh^2>1$.
Coannihilation effects, here neglected, modify the bounds
in a narrow region where $B$ and ${\ell}_R$ are nearly mass-degenerate.
If the $B$-ino freezes out during reheating the forbidden region
where $\Omega_Bh^2>1$ is a banana-shaped region.  The allowed region of
parameter space is above the charged-dark-matter region and outside
the banana-shaped region.  The size and location of the banana-shaped
disallowed region depends on $M_B/T_{RH}$.  Some examples are shown in
the figure.  
\label{figbino}} 
\end{figure}

Coannihilation effects do not significantly modify the bound on the
slepton mass for a fixed value of $M_B$ (as long as it is not too
close to ${\tilde m}_{\ell_R}$). On the other hand, these bounds can
rapidly disappear if we consider low values of the reheat
temperature. The effect is illustrated in Fig.\ \ref{figbino}, which
shows the values of slepton and $B$-ino masses incompatible with the
constraint $\Omega h^2<1$, for different choices of $T_{RH}$, in the
case of a 100\% pure $B$-ino LSP.  The upper bounds on ${\tilde
m}_{\ell_R}$ are drastically relaxed.\footnote{This is particularly
welcome in those scenarios where the supersymmetric flavor and CP
problems are avoided if the first two generations of sfermions are
heavier than a few TeV and approximately degenerate in mass.  If the
lightest supersymmetric particle is essentially $B$-ino-like then
requiring that all flavor changing neutral current and CP-violating
processes are adequately suppressed imposes a lower limit on the
$B$-ino mass of typically 200 to 300 GeV \cite{ros}.}  Therefore, this
shows that cosmological arguments based on relic abundances, used to
set upper bounds on supersymmetric particles rely on specific untested
assumptions. A low reheat temperature can completely change the
picture.

The same arguments can be applied to other supersymmetric dark matter
candidates. One possibility which one encounters in theories with
gauge-mediated supersymmetry breaking \cite{giandm,han} is given by
the messenger scalar particle with the same gauge quantum numbers of
the neutrino. The cosmological upper bound on this particle mass is
about 3 TeV, much lower than the natural theoretical expectation. A
low reheat temperature easily relaxes the bound.

The same can be said about the limits on unstable particles whose
decays into energetic products may jeopardize the successful
predictions of nucleosynthesis \cite{sarkar}. It is clear that our
results relax the bounds in the $(M_X,\tau_X)$ plane, where $\tau_X$
is the decay lifetime of the unstable particle.

\subsection{Massive  Neutrinos}  \label{nutau}

Let us now analyze the implications of our approach for massive
neutrinos, in the case in which they are stable.  First, let us
briefly recall the standard ({\it i.e.,} when the reheat temperature
is very large) prediction for the abundance of massive neutrinos.
Neutrinos are initially kept in equilibrium by weak interactions.  For
neutrinos lighter than about 1 MeV, freeze out occurs at $T_{D}\simeq
2.3$ MeV for electron neutrinos and $T_{D}\simeq 3.7$ MeV for muon and
tau neutrinos, so the neutrinos are relativistic at freeze out. The
current abundance of a generic relativistic particle can be easily
estimated to be \cite{book}
\begin{equation}
\label{st}
\Omega_X h^2 \simeq 
         \frac{g_X}{2}\, \frac{10.75}{g_*(T_D)}\, \frac{m_X}{94\,{\rm eV}}\ .
\end{equation}
Here $g_X$ is the number of degrees of freedom of the species.
Standard Model (SM) relativistic neutrinos decouple from chemical
equilibrium when $g_*(T_D)\simeq 10.75$. One usually 
concludes that the mass of SM
neutrinos cannot be larger than $94h^2$ eV, the
Cowsik--McClelland bound \cite{cw}.

If neutrino masses are larger than the freeze-out temperature, they
decouple from the thermal bath when they are nonrelativistic.  In this
case the annihilation cross section is proportional to $G_F^2m_\nu^2$,
and requiring $\Omega_\nu h^2\simlt 1$ provides an lower bound on
$m_\nu$ of about 2 GeV, the Lee--Weinberg bound \cite{lw}.  This means 
that neutrino masses in the range 94 eV $\simlt m_\nu\simlt $ 2 GeV are 
cosmologically ruled out.  This is the celebrated
Cowsik-McClelland--Lee-Weinberg (CMLW) bound \cite{cw,lw}.

This picture has to be modified in the case in which the reheat
temperature is small.  The standard CMLW bound is based on the
assumption that neutrinos have reached thermal and chemical
equilibrium in the radiation-dominated universe. This is equivalent to
the hypothesis that the maximum temperature obtained during the (last)
radiation-dominated era, that is, the reheat temperature $T_{RH}$, is
much larger than the decoupling temperature $T_D$.  We
have no physical evidence of the radiation-dominated era before the
epoch of nucleosynthesis ({\it i.e.,} temperatures above about 1 MeV).
Therefore, let us explore the possibility that the
largest temperature of the Universe during the radiation-dominated
phase is very small. Indeed, it has been recently shown that the
smallest value not excluded by nucleosynthesis at more than 95\% CL
is $T_{RH}\simeq 0.7$
MeV \cite{ka}.

\begin{figure}[t]
\centering \leavevmode\epsfxsize=300pt \epsfbox{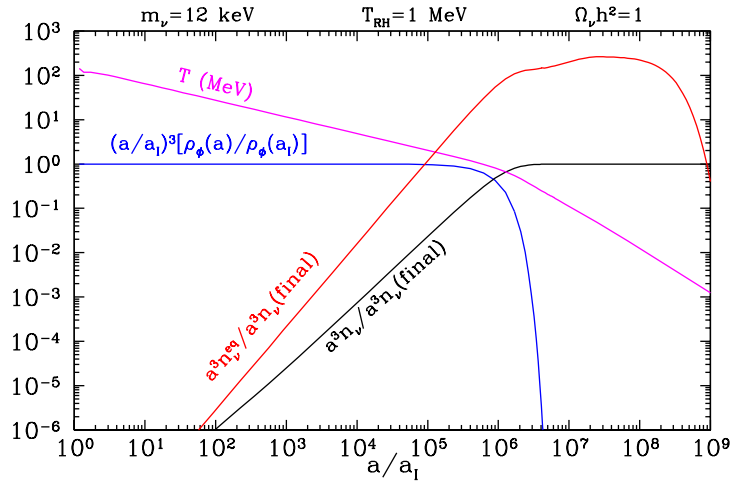}
\caption{The evolution of the number density per comoving 
volume of Dirac neutrinos for $m_\nu=12$ keV and
$T_{RH}=1$ MeV. This figure illustrates the fact that neutrinos
never attain chemical equilibrium.
\label{er_freezeout}} 
\end{figure}

Since neutrinos have only weak interactions, it is very difficult for
the thermal scatterings of particles during the reheat stage to
generate SM neutrinos through processes like $e^+ e^-\rightarrow
\nu\bar\nu$ and to bring neutrinos into chemical equilibrium. 
Furthermore, decreasing the reheat temperature increases the rate of
the expansion of the Universe, as explicitly seen in Eq.\
(\ref{hhhh}), making it more difficult for the weak interactions to
bring the neutrinos into chemical equilibrium. Therefore, if the
reheat temperature is small enough, one should expect that the SM
neutrinos produced during the reheat stage {\it never} reach chemical
equilibrium. In other words, at the beginning of the
radiation-dominated phase neutrinos populate the thermal bath, but
they have a number density, $n_\nu$, which is {\it smaller} than the
equilibrium number density, $n_\nu^{eq}$. This is well-illustrated in
Fig.\ \ref{er_freezeout}, which shows the evolution of the number
density of a muon or tau Dirac neutrino with mass 12 keV, for
$T_{RH}=1$ MeV.  It is clear that the number density only grows, and
it is always too small to catch up with the equilibrium number density,
{\it i.e.,} $n_\nu\ll n_\nu^{eq}$.  This implies that neutrino
annihilations are not efficient.

This result applies both to relativistic and nonrelativistic SM
neutrinos, and implies that the present abundance of neutrinos in low
$T_{RH}$ models is much smaller than predicted assuming that the
largest temperature of the radiation-dominated Universe was much
larger than a few MeV. This is the reason why the CMLW bound on
neutrino masses is significantly relaxed in low $T_{RH}$ models.

\begin{figure}[p]
\centering \leavevmode\epsfxsize=300pt \epsfbox{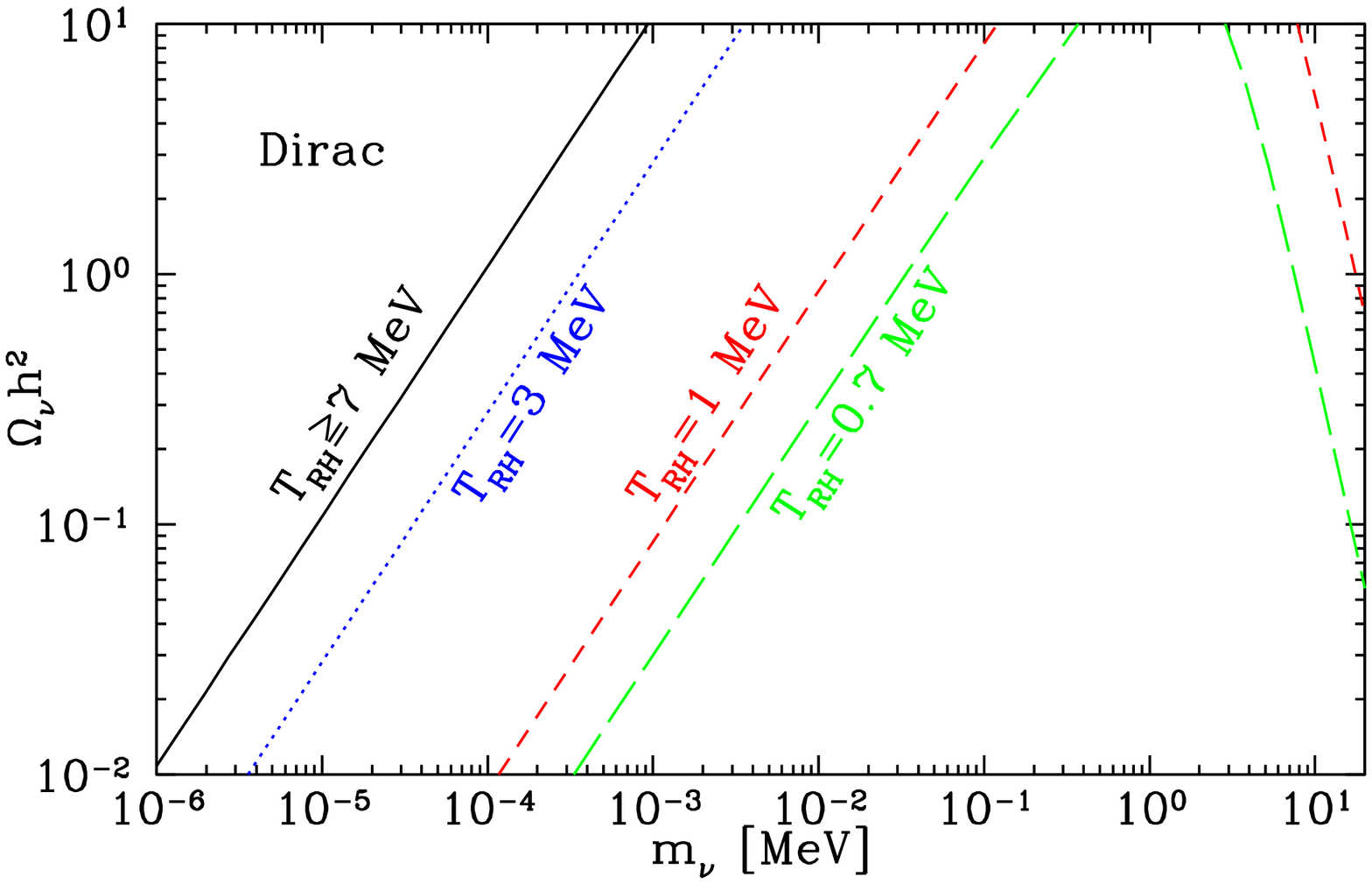}\\
\centering \leavevmode\epsfxsize=300pt \epsfbox{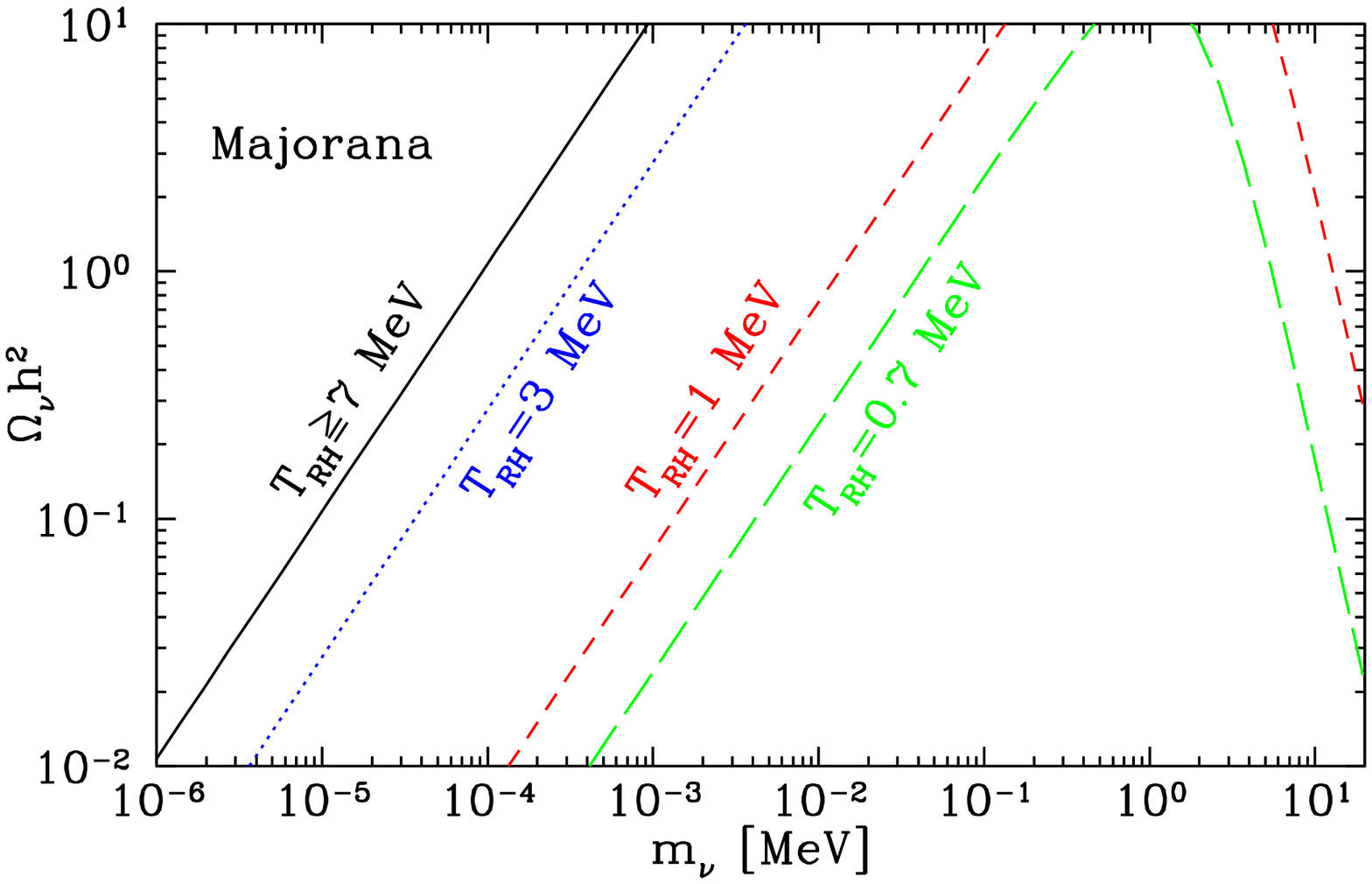}
\caption{The contribution to the closure density of Dirac and Majorana 
neutrinos for different values of the reheat temperature ($g_*=10.75$ is
used in the definition of $T_{RH}$).  
\label{neutrino}} 
\end{figure}

In Fig.\ \ref{neutrino} we present a full numerical computation of the
abundance of massive neutrinos in terms of the total $\Omega_\nu h^2$ as a
function of the mass of the neutrino, for different values of the
reheat temperature. 
In all our results we have numerically solved the
Boltzmann equation making use of the exact definition of the
thermally averaged cross section \cite{krgg}
\begin{equation}
\label{rrr}
\langle \sigma v\rangle=\frac{1}{4 m_\nu^2 T K_2^2(m_\nu/T)}
\int_{4m_\nu^2}^{\infty}\;
ds\; \sigma v E_\nu E_{\bar\nu}\sqrt{s-4 m_\nu^2} \, K_1(\sqrt{s}/T) \ ,
\end{equation}
where $s$ is the center-of-mass energy, $K_i$ are the modified Bessel
functions and we borrowed the expressions for $\sigma v E_\nu
E_{\bar\nu}$ from the appendix of Ref.\ \cite{raffelt}.\footnote{
In Ref.\ \cite{raffelt}, for the Majorana case, $C_V$ and $C_A$ have to
be interchanged. We thank G. Raffelt for communications about this
point.}

For a Dirac neutrino, the treatment of the number of degrees of freedom
requires some attention. In the relativistic case, only two of the four
helicity degrees of freedom are produced because the generation of the
wrong-helicity states is suppressed by $(m_\nu/2E_\nu)^2$.  On the other hand,
in the nonrelativistic regime, all four degrees of freedom interact with full
strength and will be produced during the reheating stage. Here, we make the
``helicity approximation'' and assume two degrees of freedom for a relativistic
neutrino species and four for a nonrelativistic neutrino species.

 At low reheat temperatures
neutrinos of a given family $\nu_\alpha$ can be produced by the
processes $e^+ e^-\rightarrow
\nu_\alpha\bar{\nu}_\alpha$ and  $\nu_\beta\bar{\nu}_\beta\rightarrow
\nu_\alpha\bar{\nu}_\alpha$, where $\beta$ is different from $\alpha$. The 
inclusion of the $e^+ e^-$ scattering in the integrated Boltzmann
equation for the number density of neutrinos $n_{\nu_\alpha}$ is
straightforward because $e^-$ and $e^+$ are kept in chemical
equilibrium by the fast electromagnetic processes and Eq.\
(\ref{binx}) applies. However, the inclusion of the
$\nu_\beta\bar{\nu}_\beta$ scatterings is more delicate because
$\nu_\beta$ and $\bar{\nu}_\beta$ are themselves not in chemical
equilibrium.   A complete solution of the problem would require a
detailed kinetic treatment of {\it all} the neutrino distribution
functions $f_\nu({\bf p},t)$ in momentum space.  This computation is
now in progress. In this paper we have limited ourselves to include
the $\nu_\beta\bar{\nu}_\beta$ scatterings by defining in
Eq.\ (\ref{rrr}) an effective cross section $\sigma_{eff}\equiv
\sigma_{e^+ e^-}+\sum_\beta (n_{\nu_\beta}/n_{\nu_\beta}^{eq})^2
\sigma_{\nu_\beta\bar{\nu}_\beta}$. To convince oneself of the validity
of this approximation, one may notice that we recover the usual
standard CMLW bound when the reheat temperature is larger than about 7
MeV. This is in agreement with the results obtained in Ref.\ \cite{ka}
where the Boltzmann equations in momentum space were numerically
solved for massless neutrinos and it was shown that for $T_{RH}\simlt$
7 MeV the effective number of massless neutrinos $N_\nu\equiv
\rho_\nu/\rho_\nu^{eq}$ starts deviating from 3.

{}From Fig.\ \ref{neutrino} we infer that in the case in which
neutrinos are relativistic, $\Omega_\nu h^2$ is approximately given by
\begin{equation}
\Omega_{\nu_\mu}h^2  = \Omega_{\nu_\tau}h^2
\simeq \left( \frac{m_\nu}{12\;{\rm keV}}
\right)\left( \frac{T_{RH}}{\rm MeV}
\right)^3\;\;\;\;\;{\rm relativistic \  Majorana \ and \ Dirac} \ .
\end{equation}

The full numerical calculation shows that neutrino
masses as large as 
$m_\nu\simeq 33$ keV are compatible with
$\Omega_\nu h^2<1$
 for the limiting
reheat temperature $T_{RH}=0.7$ MeV.  This shows that SM neutrinos
with masses up to about 33 keV are perfectly compatible with
cosmology and may even play the role of warm dark matter.

Let us now briefly see what our findings are when the neutrinos are
heavier than an MeV.  This is possible only for the tau neutrino, for
which the present experimental upper limit on the mass is 18.2 MeV
\cite{pdg}. For $m_{\nu_\tau}\sim 10$ MeV the ordinary calculation 
of the relic abundance in a radiation-dominated universe predicts a
value of $\Omega_{\nu_\tau}\sim 10^4$, which is definitely
excluded. Our findings indicate that
\begin{eqnarray}
\label{sss}
\Omega_{\nu_\tau} h^2 & = &
          \left( \frac{T_{RH}}{{\rm MeV}}
\right)^7\left( \frac{14\,{\rm MeV}}{m_{\nu_\tau}}\right)^3 
\;\;\;\;\;{\rm nonrelativistic \ Dirac} \ ,\nonumber\\
\Omega_{\nu_\tau} h^2  &=&
          \left( \frac{T_{RH}}{{\rm MeV}}
\right)^7\left( \frac{13\,{\rm MeV}}{m_{\nu_\tau}}\right)^3
\;\;\;\;\;{\rm nonrelativistic \  Majorana} \ .
\end{eqnarray}
The strong dependence upon the
reheat temperature is easily understood by realizing that for $T_{RH}$
in the MeV range we are in the condition of Sec.\ \ref{nreq}
($\alpha_s<\bar{\alpha}_s$ and $\alpha_p<\bar{\alpha}_p$), and
therefore the neutrino relic abundance is given approximately by Eq.\
(\ref{omneq}).  Our numerical result is very well reproduced
analytically by taking Eqs.\ (\ref{cross}) and (\ref{omneq}) and
using, for a Dirac neutrino,
\begin{equation}
\alpha_s = \frac{G_F^2m_{\nu_\tau}^4}{2\pi}
\left( \frac{1}{2} -2\sin^2\theta_W +4 \sin^4\theta_W \right) \ ,
\end{equation}
and, for a Majorana neutrino,
\begin{equation}
\alpha_p = \frac{2G_F^2m_{\nu_\tau}^4}{\pi}
\left( \frac{1}{2} -2\sin^2\theta_W +4
\sin^4\theta_W \right) \ .
\end{equation}
These values are obtained taking into account only the process $e^+
e^-\rightarrow \nu_\tau \bar{\nu}_\tau$ in $\sigma_{eff}$ because, as
we have numerically checked, the contribution to the production of
heavy tau-neutrinos from light neutrino annihilations is negligible at
the production time, see Eqs. (\ref{asa}) and (\ref{asb}).

It is interesting to notice that there exists a small window of
$m_{\nu_\tau}$ and $T_{RH}$ for which the ordinary tau neutrino is an
acceptable candidate for cold dark matter. However, because of the
large powers of $m_{\nu_\tau}$ and especially of $T_{RH}$ in Eq.\
(\ref{sss}), the allowed window is very limited.  
Of course in this scenario there must be a
conserved quantum number to keep the tau neutrino stable in order to
be the dark matter; this hypothesis is
in apparent contradiction with the experimental
evidence on neutrino oscillations.  

It has also been proposed that the
tau neutrino could be a cold dark matter candidate if it had a
magnetic moment of the order of $10^{-6}$ Bohr magnetons
\cite{mag}. The present experimental limit on the $\nu_\tau$ magnetic
moment of $5.4\times 10^{-7} \mu_B$ rules out this possibility. We
find that choosing $T_{RH}$ as low as 1 MeV can rescue
this possibility, since $\Omega_{\nu_\tau} h^2\simeq 1$ for a tau
neutrino magnetic moment with the maximum allowed value.

We conclude that the cosmologically disallowed region for neutrino
masses is
\begin{equation}
33\;{\rm keV}\simlt m_\nu\simlt 6\;{\rm MeV} \ 
\end{equation}
for a Dirac neutrino and
\begin{equation}
33\;{\rm keV}\simlt m_\nu\simlt 5\;{\rm MeV} \ 
\end{equation}
for a Majorana neutrino.
We stress that so long as the reheat temperature is an unknown parameter,
this should be considered the real CMLW bound on neutrino masses. Our
findings indicate that neutrinos can still play the role of warm or
cold dark matter and that the impact of massive neutrinos on
nucleosynthesis has to be revisited. These and other issues are
currently under investigation.
Finally, we note that the above values were found assuming Maxwell--Boltzmann
statistics and assuming the annihilation products are in equilibrium.

\subsection{Cosmological Bound on Axions}  \label{axionsect}

The invisible axion is still the most elegant solution to
the strong CP problem \cite{axion,reviewcp}.  An axion model has one basic
free parameter, the axion mass $m_a$, or equivalently, the Peccei-Quinn (PQ)
symmetry breaking scale $f_{PQ}$. The mass and symmetry breaking scale are 
related by
\begin{equation}
m_a\simeq 0.62 \;{\rm eV}\: \frac{10^7\; {\rm GeV}}{f_{PQ}/N} \ ,
\end{equation}
where $N$ is the color anomaly of the PQ symmetry.

Several astrophysical lower limits on $f_{PQ}$ are based on the
requirement that the axionic energy losses from stars, notably
red-giant stars, globular-cluster stars, or the core of supernova
1987A, are not in conflict with the observed properties of these
objects \cite{raf}.  These limits imply $m_a\simlt 10^{-2}$ eV
(equivalently $f_{PQ}\simgt 10^{9}$ GeV), indicating that axions, if
they exist, are both extremely light and very weakly interacting.

An upper bound on the PQ scale comes from cosmological
considerations. Let us call $\Theta$ the strong CP-violating
phase. Today $\Theta$ is anchored at the CP-conserving value,
$\Theta=0$. However, the axion mass is very temperature-dependent
\cite{mass}
\begin{equation}
\label{mt}
m_a(T)\simeq 0.1\; m_a\;\left(\frac{\Lambda_{QCD}}{T}\right)^{3.7} \ ,
\end{equation}
where $\Lambda_{QCD}\simeq 200$ MeV is the QCD scale and the relation
is valid for $T\gg \Lambda_{QCD}/\pi$.  At very high temperatures the
axion is essentially massless. This means that no special value of
$\Theta$ is specified by the dynamics, and all values of the phase are
equivalent. The axion mass turns on at a temperature $T_1$ such that
$m_a(T_1)\simeq 3 H(T_1)$, and the axion field starts evolving toward
the minimum at $\Theta=0$, eventually oscillating around it.  These
cosmic oscillations of the axion field correspond to a zero-momentum
condensate of axions which does not decay. The energy density in
axions today from this misalignment mechanism exceeds the critical
density unless $f_{PQ}\simlt 10^{12}$ GeV \cite{mis}. This result has
always been considered particularly disappointing from the theoretical
point of view, since weakly coupled string theory possesses numerous
axion candidates, whose decay constant is, however, of the order of
the string scale and therefore much larger than $10^{12}$ GeV.

The purpose of the present section is to demonstrate that the
cosmological bound on the PQ scale is significantly relaxed if we make
the assumption that the reheat temperature is smaller than the QCD
scale (similar considerations have also been made in Refs.~\cite{shafi,bd}).

Suppose then that $T_{RH}\simlt \Lambda_{QCD}\simlt T_{\rm MAX}$. This
means that the axion coherent oscillations commence when the universe
is still matter-dominated and reheating is not completed. During
this epoch the Hubble rate is given by Eq.\ (\ref{hhhh}), and the axion
mass is still given by Eq.\ (\ref{mt}) since the universe is
populated by a thermal bath between $T_{\rm MAX}$ and $T_{RH}$. Axions 
start oscillating at a temperature $T_1$ when $m_a(T_1)\simeq 3 H(T_1)$:
\begin{equation}
T_1\simeq \left(\frac{m_a}{10^{-5}\;{\rm eV}}\right)^{1/7.7}
\left[\frac{g_*(T_{RH})}{10}\right]^{1/15.4}
\left[\frac{g_*(T_{1})}{10}\right]^{-1/7.7}
\left(\frac{T_{RH}}{1\;{\rm MeV}}\right)^{2/7.7}\;160\,{\rm MeV} \ .
\end{equation}
At temperatures $T\simlt T_1$ the number density of axions scales like
$a^{-3}$ even though the axion mass is still varying.  At the
reheat temperature we have
\begin{equation}
n_a(T_{RH}) = n_a(T_1)\left[ \frac{a(T_1)}{a(T_{RH})} \right]^{3} 
= \frac{\sqrt{5\pi^3}\bar{\Theta}_1^2}{2}\left(\frac{f_{PQ}}{N}\right)^2
\frac{g_*^{3/2}(T_{RH})}{g_*(T_1)}\frac{T_{RH}^6}{T_1^4 M_{Pl}} \ ,
\end{equation}
where $\bar{\Theta}_1$ is the initial displacement of the CP-phase.
Notice, in particular, that the ratio $n_a/s$ does {\it not} remain
constant during the cosmological evolution from the temperature $T_1$
to $T_{RH}$. This is because there is a continuous release of entropy.
However, when reheating is completed and the universe enters a
radiation-dominated phase, the ratio $n_a/s$ is conserved, and one can
easily compute the present abundance of axions from the misalignment
mechanism
\begin{eqnarray}
\Omega_a h^2 & = & \frac{m_an_a(T_{RH})}{8\rho_R(T_{RH})} \, 
\frac{T_{RH}}{T_{\rm now}} \, \Omega_R h^2
\nonumber\\
&=& 2.1\times 10^{-7}
\left(\frac{\bar{\Theta}_1}{\pi/\sqrt{3}}\right)^2
\left[\frac{g_*(T_{RH})}{10}\right]^{0.24}
\left[\frac{g_*(T_{1})}{10}\right]^{-0.48}
\left(\frac{10^{-5}\;{\rm eV}}{m_a}\right)^{1.52}
\left(\frac{T_{RH}}{1\;{\rm MeV}}\right)^{1.96}.
\end{eqnarray}
Requiring that $\Omega_a h^2\simlt 1$ gives
\begin{equation}
\label{bb}
\frac{f_{PQ}}{N}\simlt 1.6\times 10^{16}
\left(\frac{T_{RH}}{1\;{\rm MeV}}\right)^{-1.3}\;{\rm GeV} \qquad 
(T_{RH}\simlt \Lambda_{QCD}\simlt T_{\rm MAX}) \ .
\end{equation}
Therefore, the cosmological axion problem is ameliorated.\footnote{In 
cosmologies where the universe is dominated early on by
the coherent oscillations of some moduli field the axion bound is
significantly weakened, as had been already observed in Ref.
\cite{shafi,bd}. In Ref. \cite{bd}, however, 
       the estimate on the upper bound on $f_{PQ}$
did not take into account the presence of the thermal bath
before the completion of reheating, and therefore neglected the
dependence of the axion mass on the temperature.}  Furthermore, in
the strong coupling vacuum described in Ref. \cite{m}, the QCD axion
might be a boundary modulus.  Dimensional analysis suggests approximately
$10^{16}$ GeV for the decay constant of such a boundary axion, not
necessarily in contradiction with the upper bound of Eq.\ (\ref{bb}),
see also the discussion in Ref.~\cite{choi}.

\section{Implications for GUT baryogenesis and leptogenesis}  
\label{baryolepto}

The explanation of the observed baryon asymmetry ($B$) in the early
universe, of the order of $10^{-11}$ in units of the entropy density,
remains a fundamental cosmological question
\cite{rt}. Several theories with typical energy scale much higher than
the electroweak scale can explain the observed baryon asymmetry. For
instance, in Grand Unified Theories (GUTs) the out-of-equilibrium
decay of heavy Higgs particles may be responsible for the direct
generation of the baryon asymmetry \cite{kt}.  Alternatively, the
baryon asymmetry may be produced from a lepton asymmetry ($L$)
\cite{fy} using the fact that any lepton asymmetry is reprocessed into
baryon number by the anomalous sphaleron transitions \cite{kl}. In the
simplest scenario, the lepton asymmetry is generated by the
out-of-equilibrium decay of a massive right-handed Majorana neutrino,
whose addition to the Standard Model spectrum breaks $B-L$.

However, any scenario for the generation of the baryon asymmetry based
on the out-of equilibrium decay of some heavy particle $X$ depends
crucially on the assumption that these particles were nearly as
abundant as photons at very high temperatures. This imposes a lower
bound on the reheat temperature, $T_{RH} \simgt M_X$. On the other
hand, in supersymmetric models the requirement that not too many
gravitinos are thermally produced after inflation provides a stringent
upper bound on the reheat temperature of about $10^{8}$ to $10^{10}$
GeV \cite{gravitino}. If this bound is violated, the decay products of
the gravitino destroy light nuclei by photodissociation and hadronic
showers, thus ruining the successful predictions of
nucleosynthesis. Therefore, any out-of-equilibrium decay scenario
would require $M_X\simlt 10^{8}$--$10^{10}$ GeV, a condition which
looks particularly problematic for GUT-inspired baryogenesis.

In order to relax this limit one usually envisages two
possibilities. Either, the heavy particles are produced directly
through the inflaton perturbative decay process \cite{infla} (this
requires that the mass of the inflaton is larger than $M_X$) or they
are generated through nonperturbative process taking place at the
preheating stage (see Ref.\ \cite{np} in the case of GUT Higgs boson
induced baryogenesis and Ref.\ \cite{np2} in the case of
leptogenesis).

In this section we wish to show that the heavy particles $X$ may be
abundantly produced by thermal scatterings during the reheat stage
even though the reheat temperature $T_{RH}$ is smaller than
$M_X$. Again, this is made possible by the fact that $T_{RH}$ is not
the maximum temperature of the universe during reheating. For the sake
of concreteness we will focus on the leptogenesis scenario, but our
findings can be easily generalized to any out-of-equilibrium scenario.

Let us indicate by $N =n_{N_1}a^3$ the number density per comoving
volume of the lightest right-handed neutrino, the one whose final
decay (into left-handed leptons and Higgs bosons) is responsible
for the generation of the lepton asymmetry.  Following the notations
of Sec.\ \ref{boltzmann}, we can write the Boltzmann equation for $N $ as
\begin{equation}
\frac{dN }{dA} =   - \frac{c_NA^{ 1/2}\left( N - N_{eq}  \right)}
{\sqrt{\Phi}} \ ,
\label{lep}
\end{equation}
where
\begin{equation}
c_N\equiv \sqrt{\frac{3}{8\pi}} \frac{M_{Pl}}{T_{RH}^2}
\left(  \Gamma_{N_1}+2\;\Gamma_{h,s}+4\; \Gamma_{h,t}\right) \ .
\end{equation}
Here $\Gamma_{N_1}$ is the decay rate of $N_1$ (for the processes
$N_1\rightarrow H^\dagger \ell_L ,H{\bar \ell}_L$); $ \Gamma_{h,s }$
and $ \Gamma_{h,t }$ are the rates of the scattering processes
containing $N_1$ in the final state, mediated by the Higgs boson in
the $s$ channel (${\bar t}_R q_L^{(3)}
\to {\bar \ell}_L N_1$) and in the $t$ channel 
($\ell_L q_L^{(3)} \to {\bar t}_R N_1$), respectively.

Let us suppose first that $T_{\rm MAX}\simlt M_1$, where $M_1$
is the $N_1$ mass. Under this
assumption we have
\begin{equation}
\label{rates}
\Gamma_{N_1} = \frac{\lambda_1^2}{8\pi} M_1 \ ,\qquad 
\Gamma_{h,s} =  \frac{3\lambda_1^2\lambda_t^2}{32 \pi^5} \;
\frac{T^3}{M_1^2}\ ,    \qquad      
\Gamma_{h,t} =  \frac{3\lambda_1^2\lambda_t^2}{32 \pi^3} \;T\;{\rm ln}
\frac{M_1}{m_h} \ ,
\end{equation}
where $\lambda_1^2\equiv (\lambda\lambda^\dagger)_{11}$, with
$\lambda_{ij}$ Yukawa coupling of $N_1$, and $\lambda_t$ is the top
Yukawa coupling.  For a more transparent physical interpretation, it
is convenient to express $\lambda_1^2$ in terms of the parameter
\begin{equation}
m_1\equiv \frac{\lambda_1^2}{2\sqrt{2}G_FM_1} \ .
\end{equation}
In the limit of small mixing angles the parameter $m_1$ coincides with
the mass of one of the light (mainly left-handed) neutrinos.

If the right-handed neutrinos do not reach an equilibrium density
($N\ll N_{eq}$), we can approximate Eq.\ (\ref{lep}) by
\begin{equation}
\frac{dN  }{dA} =    \frac{c_N A^{ 1/2}\;  N_{eq}}{\sqrt{\Phi_I      }} \ ,
\label{lep1}
\end{equation}
Along the same lines of Sec.\ \ref{nrnoneq}, we can integrate Eq.\
(\ref{lep1}) by approximating it to a Gaussian integral in the full
range between $A=1$ and $A=\infty$. This is a good approximation
because the exponential suppression in $N_{eq}$ makes the right-hand
side of Eq.\ (\ref{lep1}) negligible anywhere outside a small interval
of scale factors centered around $A=A_*$ corresponding to $T_*=M_1/10$
for the inverse decay, and to $T_*= M_1/9$ for the Higgs-mediated
$\Delta L=1$ processes. It is easy to show that the main source of
right-handed neutrinos is represented by the inverse decays whose
contribution to $N_{\infty}$ is given by
\begin{equation}
N_{\infty}\simeq \frac{c}{ g_*^{3/2}} \, 
\frac{\Gamma_{N_1} M_{Pl}^3 H_I^2 T_{RH}^3} {M_1^9} \ ,
\end{equation}
where
\begin{equation}
c=\frac{72 e^{-10} 10^9}{\sqrt{5} \pi^{11/2}}=2.7\times 10^3 \ .
\label{ceci}
\end{equation}
Notice that the final abundance is suppressed only by powers of the
right-handed mass, there is no Boltzmann suppression $\exp
(-M_1/T_{RH})$.  Furthermore, the abundance is proportional to the
rate of production (accumulation) $\Gamma_{N_1}$. However, the
consistency condition $N_{\infty}\simlt N_{eq}(T_*)$ gives an upper
bound on $\Gamma_{N_1}$
\begin{equation}
\label{bbnew}
\Gamma_{N_1}\simlt \frac{\pi}{8000}\;g_*^{1/2}\;
\frac{M_1^4}{ M_{Pl}  T_{RH}^2} \ ,
\end{equation}
or, equivalently, an upper bound on $m_1$
\begin{equation}
m_1< \left( \frac{g_*}{100} \right)^{1/2} 
\left( \frac{M_1}{T_{RH}}\right)^2 2.5\times 10^{-7}\ {\rm eV} \ .
\end{equation}
These bounds
can also be expressed in terms of a more familiar quantity
\begin{equation}
K_*\equiv \left.\frac{\Gamma_{N_1}}{H}\right|_{T=T_*}\simlt 1 \ .
\end{equation}
This condition assures that when the right-handed neutrinos are
produced, their direct decay is inefficient. The limiting case $K_*\sim
1$ would mean that the right-handed neutrinos enter into chemical
equilibrium as soon as they are generated.

The right-handed neutrinos may decay before or after the universe
reaches the reheat temperature $T_{RH}$, depending on the value of
$\lambda_1$.  Suppose that they decay after the end of the reheat
stage (which is, for instance, always the case when $M_1\simlt
8T_{RH}$). This means that at $T=T_{RH}$, the ratio of the number
density of right-handed neutrinos to the entropy density is given by
\begin{equation}
\frac{n_{N_1}}{s(T_{RH})}=\frac{25\pi  c}{g_*^{3/2}} \, 
\frac{ \Gamma_{N_1}  M_{Pl}    T_{RH}^7}{ M_1^9 } \ .
\end{equation}
This ratio remains constant until the right-handed neutrinos decay
generating a lepton asymmetry
\begin{equation}
L= \frac{10^{5}}{8}\; \, \frac{\epsilon}{g_*^{3/2} } \,
\frac{\Gamma_{N_1}  M_{Pl}     T_{RH}^7}{M_1^9} \ ,
\end{equation}
where we have indicated by $\epsilon$ the small parameter containing
the information about the CP-violating phases and the loop factors and we
have again taken into account the factor of 1/8 due to the
release of entropy after $T_{RH}$. 
The corresponding baryon asymmetry is $B=(28/79)L$, assuming only
Standard Model degrees of freedom \cite{kl}, and therefore
\begin{equation}
B=\epsilon \left(\frac{100}{g_*}\right)^{3/2} \left(\frac{T_{RH}}{M_1}
\right)^7 \left(\frac{m_1}{10^{-7}\ {\rm eV}}\right) 7\times 10^{-3} \ .
\label{cazbar}
\end{equation}
By virtue of the bound of Eq.\ (\ref{bbnew}), this baryon asymmetry is
constrained to be smaller than the critical value of $2
(\epsilon/g_*)(T_{RH}/M_1)^5$ (and of $2\times 10^{-5}\epsilon/g_*$
if we use the constraint $T_*>T_{RH}$).  
The requirement that $B$ is larger than
$2\times 10^{-11}$ implies
\begin{equation}
\label{bbbb}
M_1\simlt 16\; \left(\frac{100}{g_*}\right)^{1/5}
\left(\frac{\epsilon}{10^{-3}}\right)^{1/5} T_{RH} \ . 
\label{picchio}
\end{equation}

It is easy to convince oneself that this is also the result in the
case in which the right-handed neutrinos decay before the reheat
stage is over. Equation (\ref{picchio}) provides a necessary condition
on the mass $M_1$ of the lightest right-handed neutrino in a
leptogenesis scenario, correcting the naive estimate $M_1\simlt
T_{RH}$. The relaxation of the naive bound by more than one order of
magnitude is certainly welcome to make leptogenesis more compatible
with the cosmological gravitino problem.

Let us suppose now that $T_{\rm MAX}\simgt M_1 \simgt T_{RH} $,
and that inverse decays or production processes containing the $N_1$
in the final states can bring the right-handed neutrinos to
equilibrium before they become nonrelativistic. This amounts to
requiring that the $\Delta L=1$ interactions with total rate
$\gamma_{N_1}=\left(
\Gamma_{N_1}+2\;
\Gamma_{h,s}+4\;
\Gamma_{h,t}\right)$ 
are in thermal equilibrium at $T\simgt M_1$.  Therefore the standard
out-of-equilibrium parameter
\begin{equation}
K\equiv \left.\frac{\gamma_{N_1}}{H}\right|_{T=M_1}
\end{equation}
is larger  than unity. 

In this case, the lepton asymmetry can be written as
\begin{equation}
L=\frac{45}{2^{9/2}\pi^{7/2}}\frac{\epsilon z_f^{3/2}e^{-z_f}}{g_*}
\left( \frac{T_{RH}}{T_f}\right)^5.
\label{leppy}
\end{equation}
Here $z_f\equiv M_1/T_f$, where $T_f$ is the temperature at which the
processes that damp the baryon asymmetry go out of equilibrium, and 
the last factor in Eq.\  (\ref{leppy}) accounts for the dilution caused by
the expansion in the pre-reheat phase. We are assuming that $T_f$
is larger than the reheat temperature $T_{RH}$.

If the inverse decay dominates over scatterings (we will later quantify
this condition), then $T_f$ is approximately determined by  
\begin{equation}
\left. \Gamma_{ID}=H \right|_{T=T_f} \ ,
\label{sciap}
\end{equation}
where the inverse decay rate at $T<M_X$ is
\begin{equation}
\Gamma_{ID}=\frac{\pi^{1/2}}{2\sqrt{2}}z^{3/2}e^{-z}\Gamma_{N_1} \ .
\end{equation}
Here $z\equiv M_1/T$ and $\Gamma_{N_1}$ is given in Eq.\  (\ref{rates}).
In terms of the parameter $K$, defined by
\begin{equation}
K\equiv \left. \frac{\Gamma_{N_1}}{H} \right|_{T=M_1}=\frac{3G_Fm_1T_{RH}^2
M_{Pl}}{2\sqrt{10} \pi^{5/2}g_*^{1/2}M_1^2}=\frac{m_1}{3\times 10^{-3}\, {\rm
eV}}\left( \frac{100}{g_*} \right)^{1/2} \left( \frac{T_{RH}}{M_1}\right)^2 \ ,
\end{equation}
Eq.\  (\ref{sciap}) becomes $Kz_f^{11/2}e^{-z_f}\simeq 1$, which is
approximately solved (for $K\simlt 10^5$)
by $z_f\simeq 16 K^{0.06}$. Replacing $z_f$
in Eq.\ (\ref{leppy}), we find
\begin{equation}
L=\frac{45}{\sqrt{2} \pi^{7/2}} \left( \frac{T_{RH}}{M_1}\right)^5
\frac{\epsilon}{g_*K^{0.94}} \ .
\end{equation}
Finally, the baryon asymmetry $B=(28/79)L$ is given by
\begin{equation}
B\simeq \epsilon \left( \frac{100}{g_*} \right)^{1/2}
\left( \frac{10^{-3}\,{\rm
eV}}{m_1}\right) \left( \frac{T_{RH}}{M_1}\right)^3 6\times 10^{-3} \ .
\label{barry}
\end{equation}

Equation (\ref{barry}) is valid as long as inverse decay processes dominate
over $\Delta L=2$ scattering processes in damping the baryon asymmetry. Let
us now study the condition under which this hypothesis is justified. The
rate for lepton-violating scatterings mediated by $N_1$ exchange in the
$s$ or $t$ channels, for $T<M_1$, is given by
\begin{equation}
\Gamma_{\Delta L=2}=\frac{7G_F^2M_1^3m_1^2z^{-3}}{2\pi^3} \ .
\end{equation}
Therefore, the $\Delta L=2$ scatterings are out-of-equilibrium at high
temperatures and equilibrate at a temperature
corresponding to
\begin{equation}
z_f^{\Delta L=2}=\frac{2\sqrt{5}\pi^{9/2}g_*^{1/2}M_1}{21G_F^2M_{Pl}
m_1^2T_{RH}^2} \ .
\end{equation}
The assumption that led to Eq.\ (\ref{barry}) is then valid as long as
$z_f^{\Delta L=2}>M_1/T_{RH}$, which implies
\begin{equation}
m_1< \left( \frac{g_*}{100}\right)^{1/4}
\left( \frac{10^{10}\,{\rm GeV}}{T_{RH}}\right)^{1/2} {\rm eV} \ .
\label{scelt}
\end{equation}
When the condition (\ref{scelt}) is not satisfied, $\Delta L=2$ scatterings
lead to an exponential suppression of the baryon asymmetry.

The baryon asymmetry in Eq.\ (\ref{barry}) is valid when three
conditions are verified: the right-handed neutrinos $N_1$ reach
equilibrium, which implies $K>1$; the temperature $T_f$ is larger than
$T_{RH}$, which implies $M_1>16\;T_{RH}$; Eq.\ (\ref{scelt}) is
satisfied. These three conditions together imply a maximum value of
$M_1=(100/g_*)^{1/2} 3\times 10^{11}$\,GeV, corresponding to a maximum
$T_{RH}=(100/g_*)^{1/2} 2\times 10^{10}$\,GeV.  The maximum baryon
asymmetry, achieved when $K\simeq 1$, is $B\simeq \epsilon
(100/g_*)(T_{RH}/M_1)^5 2\times 10^{-3}$. Therefore, in this case, a
sufficiently large baryon asymmetry requires $M_1\simeq 10\; T_{RH}$
and $\epsilon \simeq 10^{-3}$.

\section{Conclusions}  \label{conclusionsect}

In this paper we studied the observational consequences of having a
reheat temperature $T_{RH}$ smaller that the characteristic
temperature at which a certain cosmological process occurs. We first
described the dynamics of reheating and derived general expressions
for the relic abundance of particles whose standard freeze-out
temperature is larger than $T_{RH}$. For nonrelativistic particles we
found two different regimes. If the annihilation cross section
$\langle \sigma v \rangle$ is smaller than the critical value in
Eqs.\ (\ref{pirl1}--\ref{pirl2}), than the present relic abundance is
proportional $\langle \sigma v \rangle$, see Eq.\ (\ref{omneq}).  In the
other case, the relic abundance is inversely proportional to $\langle
\sigma v \rangle$, see Eq.\ (\ref{oom}), but because of the fast
expansion before reheating its expression differs from the usual
result in a radiation-dominated universe.

We applied our general results on relic abundances in low-$T_{RH}$
cosmologies to several cases of interest. A first result is that the usual
unitarity bound of 340 TeV on stable-particle masses can be relaxed. The
new excluded ranges of stable-particle masses as functions of $T_{RH}$ 
are shown in Fig.\ \ref{figunitarity}.

We revisited the parameter regions of supersymmetric models
leading to viable cold dark-matter candidates in the light of low
$T_{RH}$. In particular, we found that the upper bound on the
slepton mass, as a function of the LSP $B$-ino mass, can be
significantly relaxed, as quantitatively shown in Fig.\ \ref{figbino}.
Large regions of parameter space that have been considered to be
ruled out by cosmological arguments can instead give a relic
neutralino density close to the critical value.

Next, we considered how a low reheat temperature can reduce the relic
abundance of massive neutrinos, here assumed to be stable. The
requirement that $\Omega_{\nu}h^2\simlt 1$ gives the CMLW bound
$m_{\nu}\simlt$ 94 eV only if $T_{RH}$ is larger than about 7 MeV. The
bound becomes significantly weaker for lower $T_{RH}$; for instance,
$m_{\nu}\simlt 12$ keV for $T_{RH}=1$\ MeV.
For very massive neutrinos the
Lee-Weinberg bound, $m_\nu \simgt 2$ GeV, is also modified.  Again for
$T_{RH}=1$\ MeV, the limit becomes $m_\nu >14\, (13) $ MeV for
a Dirac (Majorana) neutrino.  This means that
there is even the possibility that a stable $\nu_\tau$ with mass
consistent with the direct experimental limit ($m_{\nu_\tau} <18.2$ MeV)
freezes out when it is still nonrelativistic and becomes cold dark
matter. Indeed, for $T_{RH}=0.7$ MeV, which is the lowest value of
$T_{RH}$ not excluded by nucleosynthesis, the constraint
$\Omega_{\nu_\tau}h^2\simlt 1$ excludes only the range of masses 33
keV $<m_{\nu_\tau}<$ 6 MeV (Dirac) and 33 keV $<m_{\nu_\tau}<$ 5 MeV
(Majorana).  This result resurrects the possibility of neutrinos as
warm dark matter.  For instance, if $T_{RH}=1$ MeV, $\Omega_\nu
h^2=0.3$ for a neutrino of mass 4 keV.

The requirement that the energy stored today in the axion oscillations
(caused by a misalignment between its high-temperature and
low-temperature configurations) is not larger than the critical value
imposes a bound on the Peccei-Quinn symmetry breaking scale,
$f_{PQ}\simlt 10^{12}$ GeV.  If $T_{RH}$ is less than $\Lambda_{QCD}$,
this bound can be relaxed to values close to the GUT scale.

Finally, we have investigated the impact of low $T_{RH}$ in the
explanation of the observed baryon asymmetry by a leptogenesis
mechanism. In this context, low $T_{RH}$ means $T_{RH}<M_1$, where
$M_1$ is the mass of the lightest of the three right-handed neutrinos.
Therefore, we discuss here values of $T_{RH}$ much larger than in the
previous cases. However, the formalism is the same, because what
matters is that $T_{RH}$ is less than the relevant physical energy
scale. We have found new expressions of the baryon asymmetry as
functions of $T_{RH}$, see Eqs.\ (\ref{cazbar}) and (\ref{barry}). 
Moreover, values of
$M_1$ an order of magnitude larger than $T_{RH}$ can still lead to a
sufficient density of right-handed neutrinos to explain the baryon
asymmetry. This is a welcome and important result, when one tries to
make leptogenesis consistent with the cosmological gravitino problem.

\acknowledgements{
We have benefitted from conversations with N. Arkani-Hamed,
R. Barbieri, M. Barnett, S. Dodelson, A. Dolgov, L. Hall, H. Murayama,
G. Raffelt, R. Rattazzi, and L. Roszkowski.  
AR thanks the Theoretical Astrophysics
Group at Fermilab, the Theory Group at NYU and the LBNL Theory Group
for the kind hospitality during the last stages of this work.  EWK was
supported by the Department of Energy and NASA under Grant NAG5-7092.}

\setcounter{equation}{0}
\renewcommand{\theequation}{A\arabic{equation}}
\section*{APPENDIX: EVOLUTION AFTER $T_{RH}$}
Numerical results show that there is about a factor of 8 increase in the
comoving entropy after $T_{RH}$, and that the value of $\Phi$ at $T=T_{RH}$
is
$\Phi_{RH}=0.79 \Phi_I$, {\it i.e.,} only about 21\% of the comoving $\phi$
energy density has been extracted at reheat time.
In this Appendix we will give an explanation of these results.

Assume that at $T_{RH}$ the radiation energy density is less than the $\phi$
energy density.  Then the evolution equation for $\Phi$ is
\begin{equation}
\frac{d\Phi}{dA}=-\left(\frac{\pi^2g_*}{30}\right)^{1/2} A^{1/2} \Phi^{1/2} \ .
\end{equation}
Integrating this equation from $A=1$
to $A=A_{RH}$ with the initial condition $\Phi=\Phi_I$,  we obtain
\begin{equation}
\left( \frac{\Phi_{RH}}{\Phi_I} \right)^{1/2} =1   
-\frac{1}{3}\left(\frac{\pi^2g_*}{30}\right)^{1/2} \Phi_I^{-1/2}
A_{RH}^{3/2} \qquad (A_{RH}\gg 1) \ .
\end{equation}
Evaluating Eq.\ (\ref{temp}) at $T=T_{RH}$ we obtain
\begin{equation}
A_{RH}^3= \frac{24}{5\pi^2g_*} \Phi_I .
\label{piffetepaf}
\end{equation}
Using this relation, we can estimate 
\begin{equation}
\frac{\Phi_{RH}}{\Phi_I} = \left(\frac{13}{15}\right)^2 \sim 0.75 \ .
\end{equation}
This is in good agreement with the numerical results 
and is independent of model parameters.

The second step is an estimate of how much entropy is released after
$T_{RH}$, if $\Phi_{RH}=0.75 \Phi_I$.
Let us make the crude approximation that the entropy release is instantaneous 
just after $T_{RH}$.  Then $\Delta\rho_R = \Delta\rho_\phi$, and therefore
\begin{equation}
\Delta \rho_R = 
0.75~\frac{\Phi_I T_{RH}^4}{A_{RH}^3} \ .
\end{equation}
Using Eq.\ (\ref{piffetepaf}) we obtain
\begin{equation}
\Delta \rho_R = 0.16~\pi^2g_* T_{RH}^4,
\end{equation}
which gives
\begin{equation}
\frac{\Delta \rho_R}{\rho_R} \simeq 5.
\end{equation}

Therefore the analytic estimate in the case of the instantaneous
approximation 
predicts a model-independent factor of 
$5^{3/4}\simeq 3.4$ for the release of entropy after $T_{RH}$.  
The model dependence 
comes in if we relax the assumption of instantaneous release of entropy,
increasing  the estimate,  since $\rho_\phi$ redshifts like $A^{-3}$.
The numerical result shows an increase of about a factor of 8 in the
comoving entropy after $T_{RH}$.


\end{document}